\documentclass[showpacs,amsmath,amssymb]{revtex4}
\usepackage{graphicx}
\usepackage{dcolumn}
\usepackage{bm}
\newcommand{\be}{\begin{eqnarray}}
\newcommand{\en}{\end{eqnarray}}
\newcommand{\ben}{\begin{eqnarray*}}
\newcommand{\enn}{\end{eqnarray*}}
\newcommand{\pa}{\partial}
\newcommand{\na}{\nabla}
\newcommand{\f}{\frac}
\newcommand{\m}[1]{\mathbb{#1}}

\newcommand{\p}{\paragraph{}}
\newcommand{\bi}{\begin{itemize}}
\newcommand{\ei}{\end{itemize}}

\newcommand{\la}{\langle}
\newcommand{\ra}{\rangle}

\newcommand{\R}{\Rightarrow}
\renewcommand{\r}{\rho}
\renewcommand{\p}{\bot}
\renewcommand{\a}{\alpha}
\renewcommand{\b}{\beta}
\newcommand{\mega}{Q}
\renewcommand{\O}{\Omega}
\renewcommand{\p}{\bot}
\newcommand{\td}{\tilde}

\begin{document}
%%%%%%%%%%%%%%%%%%%%%%%%%%%%%%%%%%%%%%%%%%%%%%%%%%%%%%%%%%%%%%%%%%%%%%%%%%%%%%%%%%%%%%%%%%%%%%%%%%%%%%%%%%%%%%%%%%%%%%%%%%%%
\title{Role of Third-Order Structure Function in Studying Two-Dimensionalisation of Turbulence}
%%%%%%%%%%%%%%%%%%%%%%%%%%%%%%%%%%%%%%%%%%%%%%%%%%%%%%%%%%%%%%%%%%%%%%%%%%%%%%%%%%%%%%%%%%%%%%%%%%%%%%%%%%%%%%%%%%%%%%%%%%%%
\author{Sagar Chakraborty}
\email{sagar@bose.res.in}
\affiliation{S.N. Bose National Centre for Basic Sciences\\Saltlake, Kolkata 700098, India}
%%%%%%%%%%%%%%%%%%%%%%%%%%%%%%%%%%%%%%%%%%%%%%%%%%%%%%%%%%%%%%%%%%%%%%%%%%%%%%%%%%%%%%%%%%%%%%%%%%%%%%%%%%%%%%%%%%%%%%%%%%%%
\date{\today}
%%%%%%%%%%%%%%%%%%%%%%%%%%%%%%%%%%%%%%%%%%%%%%%%%%%%%%%%%%%%%%%%%%%%%%%%%%%%%%%%%%%%%%%%%%%%%%%%%%%%%%%%%%%%%%%%%%%%%%%%%%%%
\begin{abstract}
We look at various correlation functions, which include those that involve both the velocity and the vorticity fields, in two-dimensional (2D) isotropic homogeneous unforced turbulence.
We adopt the more intuitive approach due to Kolmogorov (and subsequently, Landau in his text on fluid dynamics) and show that how the 2D turbulence's results, obtainable using other methods, may be established in a simpler way.
Same method is used to calculate some third-order structure functions for quasi-geostrophic (QG) turbulence for the forward cascade of pseudo-potential enstrophy and the inverse energy cascade in quasi-geostrophic turbulence.
These results motivate us to study the two-point third order structure function in the context of the two-dimensionalisation effect.
Consequent studies enable us to give a reason for the inverse energy cascade in the two-dimensionalised rapidly rotating three dimensional (3D) incompressible turbulence.
For such a system, literature shows a possibility of the exponent of wavenumber in the energy spectrum's relation to lie between -2 and -3.
We argue the existence of a stricter range of -2 to -7/3 for the exponent in the case of rapidly rotating turbulence which is in accordance with the recent experiments.
Also, a derivation for the two point third order structure function has been provided helping one to argue that even with slow rotation one gets, although dominated, a spectrum with the exponent -2.87, thereby hinting at the initiation of the two-dimensionalisation effect with rotation.
Moreover, using the Gledzer-Ohkitani-Yamada (GOY) shell model, modified for rotation, these signatures of two-dimensionalisation effect have been verified.

\end{abstract}
%%%%%%%%%%%%%%%%%%%%%%%%%%%%%%%%%%%%%%%%%%%%%%%%%%%%%%%%%%%%%%%%%%%%%%%%%%%%%%%%%%%%%%%%%%%%%%%%%%%%%%%%%%%%%%%%%%%%%%%%%%%%
%%%%%%%%%%%%%%%%%%%%%%%%%%%%%%%%%%%%%%%%%%%%%%%%%%%%%%%%%%%%%%%%%%%%%%%%%%%%%%%%%%%%%%%%%%%%%%%%%%%%%%%%%%%%%%%%%%%%%%%%%%%%
\maketitle
%%%%%%%%%%%%%%%%%%%%%%%%%%%%%%%%%%%%%%%%%%%%%%%%%%%%%%%%%%%%%%%%%%%%%%%%%%%%%%%%%%%%%%%%%%%%%%%%%%%%%%%%%%%%%%%%%%%%%%%%%%%%
\section{Introduction}
Rotation, in the face of the discovery of two-dimensionalisation effect, has emerged as a parameter that can progressively make a 3D turbulent flow look like a quasi-2D or a 2D turbulent flow.
The phrase `look like' basically means that certain properties of 3D turbulence, such as wavenumber dependence of energy spectrum, direction of energy cascade {\it etc.}, become such that they give impression that the flow is getting two-dimensionalised.
In view of the fact that the dynamics of oceans, atmospheres, liquid planetary cores, fluid envelopes of stars and, other bodies of astrophysical and geophysical interest do require an understanding of inherent properties of turbulence in the rotating frame of reference, the problem of two-dimensionalisation is of central interest to any serious scientist; turbulence in rotating bodies is even of some industrial and engineering interest.
\\
In the steady non-turbulent flow, for low Rossby number ($Ro=U/2L\O$) and high Reynolds number ($Re=UL/\nu$), Taylor-Proudman theorem\cite{Batchelor} argues that rotation two-dimensionalises the flow.
This argument is often carelessly extended to turbulent flows to explain the rotation induced two-dimensionalisation arising therein.
The two-dimensionalisation of the 3D turbulent flow in presence of rotation has begun to be understood as a subtle non-linear effect which is distinctly different from Taylor-Proudman effect.
\\
Cambon {\it et al.}\cite{Cambon1} have showed that in the presence of rotation, the transfer of energy from small to high wavenumbers is inhibited; at the same time, the strong angular dependence of this effect leads to a draining of the spectral energy from the parallel to the normal wave vectors (w.r.t. the rotation axis) showing a trend towards two-dimensionalisation.
Waleffe\cite{Waleffe} has used helical decomposition of the velocity field to study the nature of triad interactions in homogeneous turbulence and coupling it with the instability assumption predicted a transfer of energy toward wave vectors perpendicular to the rotation axis under rapid rotation.
The helical decomposition turns out to be very handy to deal with rapidly rotating turbulent flow.
In that case the linear eigensolutions of the problem, the so-called inertial waves, have the structure of helical modes.
The assumption about the triadic transfers, coupled with resonance condition for non-linear interaction between inertial waves, show that there will be a tendency toward non-linear two-dimensionalisation of the flow.
\\
Simulations by Smith {\it et al.}\cite{Smith} speak volumes for the two-dimensionalisation effect.
They show the coexistence of inverse cascade (a typical feature of 2D turbulence) and forward cascade in forced rotating turbulence within a periodic box of small aspect ratio.
In the simulations, the ratio of the mean rates of energy dissipated to the energy injected decreased almost linearly, for $Ro$ less than a critical value, with decrease in $Ro$ (increase in angular velocity $|\vec{\O}|$).
By the way, a very recent numerical study\cite{Waite} shows similar transition from stratified to quasi-geostrophic turbulence, manifested by the emergence of an inverse cascade -- a conclusion that agrees with that of Lindborg\cite{LindborgQG}.
\\
Although recent experiments by Baroud {\it et al.}\cite{Baroud1,Baroud2} and Morize {\it et al.}\cite{Morize1,Morize2}  have shed some light on the two-dimensionalisation effect, the scaling of two-point statistics and energy spectrum in rotating turbulence remains a controversial topic.
Zhou\cite{Zhou} in analogy with MHD turbulence has proposed an energy spectrum $E(k)\sim k^{-2}$ for rapidly rotating 3D turbulent fluid (also see \cite{Canuto}) and this does seem to be validated by some experiments\cite{Baroud1,Baroud2} and numerical simulations\cite{Yeung,Hattori,Reshetnyak,Muller}.
But some experiments\cite{Morize1} do not tally with this proposed spectrum.
They predict steeper than $k^{-2}$ spectrum and this again seem to be drawing some support from numerical results\cite{Yang,Bellet} and analytical results found using wave turbulence theory\cite{Galtier,Cambon2}.
\\
Unbiasedly speaking, if one wishes angular velocity to become a relevant parameter in constructing the energy spectrum $E(k)$, simple dimensional analysis would lead one to:
\be
E(k)\propto\O^{\frac{3m-5}{2}}\varepsilon^{\f{3-m}{2}}k^{-m}
\label{0}
\en
where $m$ is a real number.
$m$ should be restricted within the range 5/3 to 3 to keep the exponents of $\O$ and $\varepsilon$ in relation (\ref{0}) positive.
The two limits $m=5/3$ and $m=3$ corresponds to isotropic homogeneous 3D turbulence and 2D turbulence respectively.
The spectrum due to Zhou --- $E(k)\sim k^{-2}$ --- is due an intermediate value of $m=2$.
So, as far as the present state of the literature on rotating turbulence goes, two-dimensionalisation of 3D turbulence would mean the dominance of a spectrum which goes towards $E(k)\sim k^{-3}$ and which may choose to settle at $E(k)\sim k^{-2}$ --- an issue yet to be fully resolved.
\\
As a turbulent flow can be treated as the manifestation of a random velocity field, one actually hopes to unveil the statistical properties of the flow rather than every other detail of the flow.
This means that we basically are after some probability distribution for the flow: Even today, this remains a tough nut to crack.
However, the knowledge of the structure functions assists one to take the first step towards finding the distribution.
The structure functions are experimentally measurable and hence are of extreme practical importance.
The scaling relations of the structure functions, thus, are the lynchpins of turbulence theory although uncertainty lingers as to their general validity and the details of the derivation as far as the present status of research in turbulence is concerned.
Therefore, naturally a lot of time and effort are spent by the scientists working in the field of turbulence to determine the exact forms for these functions and to study various phenomena in their light.
\\
In this article, we shall deal with the phenomenon of two-dimensionalisation of 3D incompressible high Reynold's number fluid turbulence and try to see what can be said about it from the study of structure functions, especially $S_3$ (to be defined below).
Basically, herein we shall comprehensively review the works\cite{SagarE0, SagarE5, SagarE4, SagarE1, SagarE2, SagarE3} done in this direction by the author and observe how the method of calculating structure functions, developed by Kolmogorov and subsequently Landau, serves as the cornerstone for studying the two-dimensionalisation effect from the angle adopted by the author.
\section{2D turbulence}
It may be unanimously accepted that Kolmogorov's four-fifths law\cite{Kolmogorov} is a landmark in the theory of turbulence because it is an exact non-trivial result.
In three spatial dimensions, this law says that the two-point third order velocity correlation function behaves as:
\ben
S_3\equiv\left<\left[\left\{\vec{v}(\vec{x}+\vec{l})-\vec{v}(\vec{x})\right\}.\f{\vec{l}}{|\vec{l}|}\right]^3\right>=-\f{4}{5}\varepsilon l
\enn
where $\varepsilon$ is the rate per unit mass at which energy is being transferred through the inertial range.
$\vec{v}$ is the velocity field.
The inertial range is the intermediate spatial region postulated by Kolmogorov where the large scale disturbances (flow maintaining mechanisms) and the molecular scale viscous dissipation play no part.
This result is of such central significance that attempts are regularly made to understand it afresh and to extend it in other situations involving turbulence.
There appears to be following two primary methods\cite{Bhattacharjee} of obtaining this result:
\begin{enumerate}
\item
The original Kolmogorov method put forward in details in the fluid dynamics text due to Landau and Lifshitz\cite{Landau}.
There is no external forcing in this approach, and the equality of dissipation rate and forcing rate for the energy is never enforced.
\item
A field-theoretic technique that invokes the so-called `dissipation anomaly' in the high Reynold's number fluid turbulence.
In this approach, there is an external forcing that maintains a steady state turbulence.
\end{enumerate}
The two approaches yield the same important results as they should.
\\
If one goes by the standard procedure given in the book by Frisch\cite{Frisch} to derive the form of the correlation function in $d$-D turbulence with the {assumption} of forward energy cascade, one would land up on\cite{Gawedzki}:
\be
S_3=-\f{12}{d(d+2)}\varepsilon l
\label{rev1}
\en
where $\varepsilon$ is the mean rate of dissipation of energy per unit mass.
This result is not quite true for the two-dimensional case since it gives for $d=2$: $S_3=-(3/2)\varepsilon l$ and not $S_3=(3/2)\varepsilon l$.
This is so because the calculation doesn't take into account the conservation of enstrophy in 2D turbulence which causes the reverse cascade of energy\cite{Kraichnann}.
It might be noted that $S_3=(3/2)\varepsilon l$ for $d=2$ is for the regime of scales larger than the forcing scale\cite{SagarE0,Bernard}.
\\
Actually, if we consider the two-dimensional turbulence, then in the inviscid limit, we have two conserved quantities -- energy and enstrophy.
This gives rise to two fluxes with the enstrophy flux occurring from the larger to the smaller spatial scales.
The energy flux goes in the reverse direction.
Recently, Bernard\cite{Bernard} and Lindborg\cite{Lindborg2D} have used the above mentioned techniques to obtain the third order structure function for both the energy and the enstrophy cascade regions in forced 2D turbulence.
We believe that the issue is important enough that a derivation of these results using the Kolmogorov-Landau approach should be useful.
This is what we have attempted here and our results do come out in agreement with them.
Besides, we also have derived some other correlation functions which deal with vorticity fields in the inertial region and also some two-point second order correlation functions in the dissipative region following the arguments of Landau, thereby consolidating the equivalence between the two approaches mentioned in the beginning.
\subsection{Second order velocity correlation function}
It is a well-established fact that there exists a direct-cascade of enstrophy in 2D turbulence.
One defines total enstrophy as $\Gamma=\f{1}{2}\int_{\textrm{all space}}\omega^2 d^2\vec{\r}$ where $\omega=\pa_xv_y-\pa_yv_x$ is the vorticity in the Cartesian coordinates; $\vec{v}$ being the velocity field.
As we shall consider incompressible fluids only ($\vec{\nabla}.\vec{v}=0$), we shall take density to be unity and let $\vec{\r}$ take over the task of representing position vector in 2D plane.
The enstrophy flows through the inertial range and gets dissipated near dissipation scale.
Using the antisymmetric symbol $\varepsilon_{\a\b}$ that has four components, viz. $\varepsilon_{11}=\varepsilon_{22}=0$ and $\varepsilon_{12}=-\varepsilon_{21}=1$, one may define the mean rate of dissipation of enstrophy per unit mass as:
\be
&{}&\eta\equiv\nu\la \vec{\nabla}\omega.\vec{\nabla}\omega\ra\\
\Rightarrow&{}&\eta=\nu\varepsilon_{\tau\a}\varepsilon_{\theta\b}\la (\pa_\tau\pa_\gamma v_{\a})(\pa_\theta\pa_\gamma v_{\b})\ra
\label{cr10}
\en
Here, angular brackets denote an averaging procedure which averages over all possible positions of points $1$ and $2$ at a given instant of time and a given separation.
Now, if $\vec{v}_1$ and $\vec{v}_2$ represent the fluid velocities at the two neighbouring points at $\vec{\r}_1$ and $\vec{\r}_2$ respectively, one may define rank two correlation tensor:
\be
B_{\a\b}\equiv\la(v_{2\a}-v_{1\a})(v_{2\b}-v_{1\b})\ra
\label{cr11}
\en
For simplicity, we shall take a rather idealised situation of turbulence flow which is homogeneous and isotropic on every scale --- a case achievable in practice in a vigorously-shaken-fluid left to itself.
The component of the correlation tensor will obviously, then, be dependent on time, a fact which won't be shown explicitly in what follows.
As the features of local turbulence is independent of averaged flow, the result derived below is applicable also to the local turbulence at a distance $\r$ much smaller than the fundamental scale.
Isotropy and homogeneity suggests following general form for $B_{\a\b}$
\be
B_{\a\b}=A_1(\r)\delta_{\a\b}+A_2(\r)\r^o_\a\r^o_\b
\label{cr12}
\en
where $A_1$ and $A_2$ are functions of time and $\r$.
The Greek subscripts can take two values $\r$ and $\p$ which respectively mean the component along the radial vector $\r$ and the component in the transverse direction.
Einstein's summation convention will be used extensively.
Also,
\be
\vec{\r}=\vec{\r}_2-\vec{\r}_1,\phantom{xxx}\r^o_\a\equiv\r_\a/{|\vec{\r}|},\phantom{xxx}\r^o_\r=1,\phantom{xxx}\r^o_\p=0
\en
using which in the relation (\ref{cr12}), one gets:
\be
B_{\a\b}=B_{\p\p}(\delta_{\a\b}-\r^o_\a\r^o_\b)+B_{\r\r}\r^o_\a\r^o_\b
\label{cr13}
\en
One may break the relation (\ref{cr11}) as
\be
B_{\a\b}=\la v_{1\a}v_{1\b}\ra+\la v_{2\a}v_{2\b}\ra- \la v_{1\a}v_{2\b}\ra-\la v_{2\a}v_{1\b}\ra
\label{cr14}
\en
and defining $b_{\a\b}\equiv\la v_{1\a}v_{2\b}\ra$, one may proceed, keeping in mind the isotropy and the homogeneity, to write
\be
B_{\a\b}=\la v^2\ra\delta_{\a\b}-2b_{\a\b}
\label{cr16}
\en
Again, having assumed incompressibility, one may write:
\be
&{}&\pa_\b B_{\a\b}=0\\
\Rightarrow&{}&B_{\p\p}=\r B'_{\r\r}+B_{\r\r}
\label{cr17}
\en
where the equation (\ref{cr13}) has been used and prime ($'$) is allowed to denote derivative w.r.t. $\r$.
Near the dissipation region the flow is regular and its velocity varies smoothly which allows to expand $v$ in a series of power of $\r$.
One must take $v\sim \r^2$ neglecting the higher powers ($v\sim \r$ is not taken because it leads to the contradictory result that $\eta=0$ as can be seen from the relation (\ref{cr10})).
So, treating $a$ as a proportionality constant, let $B_{\r\r}=a\r^4$, which means $B_{\p\p}=5a\r^4$ (using equation (\ref{cr17})) and hence,
\be
&{}&\la v_{1\a}v_{2\b}\ra=\f{1}{2}\la v^2\ra\delta_{\a\b}-\f{5}{2}a\r^4\delta_{\a\b}+2a\r^2\r_\a\r_\b\\
\Rightarrow&{}&\la (\pa_{1\tau}\pa_{1\gamma}v_{1\a})(\pa_{2\theta}\pa_{2\gamma}v_{2\b})\ra=-72a\delta_{\theta\tau}\delta_{\a\b}+24a\delta_{\b\theta}\delta_{\a\tau}+24a\delta_{\a\theta}\delta_{\b\tau}\\
\Rightarrow&{}&\varepsilon_{\tau\a}\varepsilon_{\theta\b}\la (\pa_\tau\pa_\gamma v_{\a})(\pa_\theta\pa_\gamma v_{\b})\ra=-192a\\
\label{cr18-cr19}
\Rightarrow&{}&B_{\r\r}=-\f{\eta\r^4}{192\nu}
\label{cr19}
\en
In the equation (\ref{cr18-cr19}), we have put $\vec{\r}_1\approx\vec{\r}_2$, for these relations are assumed to be valid for arbitrarily small $\r$.
While writing the relation (\ref{cr19}), equation (\ref{cr10}) has been recalled.
This $B_{\r\r}$ is the two-point second order correlation function for enstrophy cascade in dissipation range.
\subsection{Third order velocity correlation function}
Now, we shall focus thoroughly on the inertial range.
Let's again define:
\ben
b_{\a\b,\gamma}\equiv\la v_{1\a}v_{1\b}v_{2\gamma}\ra
\enn
Invoking homogeneity and isotropy once again along with the symmetry in the first pair of indices, one may write the most general form of the third rank Cartesian tensor for $b_{\a\b,\gamma}$ as
\be
b_{\a\b,\gamma}&=&C(\r)\delta_{\a\b}\r^o_\gamma+D(\r)(\delta_{\gamma\b}\r^o_\a+\delta_{\a\gamma}\r^o_\b)+F(\r)\r^o_\a\r^o_\b\r^o_\gamma
\label{cr110}
\en
where, $C$, $D$ and $F$ are functions of $\r$.
Yet again, incompressibility dictates:
\be
\f{\pa}{\pa\rho_{2\gamma}}b_{\a\b,\gamma}=\f{\pa}{\pa\rho_{\gamma}}b_{\a\b,\gamma}=0\\
\Rightarrow C'\delta_{\a\b}+\f{C}{\r}\delta_{\a\b}+\f{2D}{\r}\delta_{\a\b}+\f{2D'}{\r^2}\r_\a\r_\b-\f{2D}{\r^3}\r_\a\r_\b+\f{F'}{\r^2}\r_\a\r_\b+\f{F}{\r^3}\r_\a\r_\b=0
\label{cr111}
\en
Putting $\a=\b$ in equation (\ref{cr111}) one gets:
\be
2C+2D+F=\f{\textrm{constant}}{\r}=0
\label{cr111-cr112}
\en
where, it as been imposed that $b_{\a\b,\gamma}$ should remain finite for $\r=0$.
Again, using equation (\ref{cr111}), putting $\a\ne\b$ and manipulating a bit one gets:
\be
D=-\f{1}{2}(\r C'+C)
\label{cr112}
\en
using which in relation (\ref{cr111-cr112}), one arrives at the following expression for $F$:
\be
F=\r C'-C
\label{cr113}
\en
Defining
\be
B_{\a\b\gamma}&\equiv&\la(v_{2\a}-v_{1\a})(v_{2\b}-v_{1\b})(v_{2\gamma}-v_{1\gamma})\ra\nonumber\\
&=&2(b_{\a\b,\gamma}+b_{\gamma\b,\a}+b_{\a\gamma,\b})
\label{cr113-cr114}
\en
and putting relations (\ref{cr112}) and (\ref{cr113}) in the equation (\ref{cr113-cr114}) and using relation (\ref{cr110}), one gets:
\be
&{}&B_{\a\b\gamma}=-2\r C'(\delta_{\a\b}\r^o_\gamma+\delta_{\gamma\b}\r^o_\a+\delta_{\a\gamma}\r^o_\b)+6(\r C'-C)\r^o_\a\r^o_\b\r^o_\gamma\\
\label{cr114}
\Rightarrow&{}&S_3\equiv B_{\r\r\r}=-6C
\label{cr115}
\en
which along with relations (\ref{cr112}), (\ref{cr113}) and (\ref{cr110}) yields the following expression:
\be
&{}&b_{\a\b,\gamma}=-\f{S_3}{6}\delta_{\a\b}\r^o_\gamma+\f{1}{12}(\r S'_3+S_3)(\delta_{\gamma\b}\r^o_\a+\delta_{\a\gamma}\r^o_\b)-\f{1}{6}(\r S'_3-S_3)\r^o_\a\r^o_\b\r^o_\gamma
\label{cr116}
\en
Navier-Stokes equation suggests:
\be
\f{\pa}{\pa t}v_{1\a}=-v_{1\gamma}\pa_{1\gamma}v_{1\a}-\pa_{1\a}p_1+\nu\pa_{1\gamma}\pa_{1\gamma}v_{1\a}
\label{cr117}\\
\f{\pa}{\pa t}v_{2\b}=-v_{2\gamma}\pa_{2\gamma}v_{2\b}-\pa_{2\b}p_2+\nu\pa_{2\gamma}\pa_{2\gamma}v_{2\b}
\label{cr118}
\en
multiplying equations (\ref{cr117}) and (\ref{cr118}) with $v_{2\b}$ and $v_{1\a}$ respectively and adding subsequently, one gets the following after averaging the consequent result:
\be
\f{\pa}{\pa t}\la v_{1\a}v_{2\b}\ra&=&-\pa_{1\gamma}\la v_{1\gamma}v_{1\a}v_{2\b}\ra-\pa_{2\gamma}\la v_{2\gamma}v_{1\a}v_{2\b}\ra\nonumber\\
&{}&-\pa_{1\a}\la p_1v_{2\b}\ra-\pa_{2\b}\la p_2v_{1\a}\ra\ra\nonumber\\
&{}&+\nu \pa_{1\gamma}\pa_{1\gamma}\la v_{1\a}v_{2\b}+\nu \pa_{2\gamma}\pa_{2\gamma}\la v_{1\a}v_{2\b}\ra
\label{cr119}
\en
Due to isotropy, the correlation function for the pressure and velocity, ($\la p_1\vec{v}_2\ra$), should have the form $f(\r)\vec{\r}/|\vec{\r}|$.
But since $\pa_{\a}\la p_1{v}_{2\a}\ra=0$ due to solenoidal velocity field, $f(\r)\vec{\r}/|\vec{\r}|$ must have the form $\textrm{constant}\times(\vec{\r}/|\vec{\r}|^2)$ that in turn must vanish to keep correlation functions finite even at $\r=0$.
Thus, equation (\ref{cr119}) can be written as:
\be
\f{\pa}{\pa t}b_{\a\b}=\pa_{\gamma}(b_{\a\gamma,\b}+b_{\b\gamma,\a})+2\nu \pa_\gamma\pa_\gamma b_{\a\b}
\label{cr120}
\en
For isotropic and homogeneous turbulence, the condition of incompressibility gives the easily derivable well-known result:
\be
4\pa_\gamma b_{\a\gamma,\a}=\pa_\gamma B_{\a\a\gamma}
\label{neww1}
\en
Defining $W\equiv\la\omega_1\omega_2\ra$ and noting that $W=-\pa_\delta\pa_\delta b_{\a\a}$, we get from relations (\ref{cr120}) and (\ref{neww1}):
\be
-\f{\pa W}{\pa t}=\f{1}{2}\pa_\delta\pa_\delta(\pa_{\gamma}B_{\a\a\gamma})-2\nu\pa_\delta\pa_\delta W
\label{neww2}
\en
Again, if one defines $\Omega\equiv\la(\omega_2-\omega_1)(\omega_2-\omega_1)\ra$ (which is not to be confused with the rotation rate discussed earlier), for homogeneous isotropic turbulence one may write $\Omega=2\la\omega^2\ra-2W$.
So, equation (\ref{neww2}) can be manipulated into the following:
\be
&&\f{1}{2}\f{\pa \Omega}{\pa t}-\f{\pa \la\omega^2\ra}{\pa t}=\f{1}{2}\pa_\delta\pa_\delta\pa_{\gamma}B_{\a\a\gamma}+\nu\pa_\delta\pa_\delta \Omega-2\nu\pa_\delta\pa_\delta\la\omega^2\ra\\
\Rightarrow&&\pa_\delta\pa_\delta\pa_{\gamma}B_{\a\a\gamma}=4\eta\\
\Rightarrow&&B_{\a\a\rho}=\f{1}{4}\eta\rho^3\label{neww3}
\en
Here, we have assumed $\f{\pa \Omega}{\pa t}$ to be relatively negligible and let $\nu\rightarrow 0$ so that the terms proportional to $\nu$ vanish.
Also, we have recalled that $\f{1}{2}\f{\pa \la\omega^2\ra}{\pa t}=-\eta$.
From equations (\ref{cr110}) and (\ref{cr113-cr114}), and the condition of incompressibility, it readily follows that $B_{\bot\bot\rho}=\f{\rho}{3}\f{\pa}{\pa\rho}B_{\r\r\r}$ putting which in expression (\ref{neww3}) and integrating subsequently (keeping in mind that $B_{\r\r\r}$ shouldn't blow up at $\rho=0$), we arrive at:
\be
B_{\r\r\r}=+\f{1}{8}\eta\r^3\label{neww4}
\en
This is the one-eighth law for the unforced 2D incompressible turbulence proved using the Kolmogorov-Landau approach.
\\
Let us go back to equation (\ref{cr120}).
Using expressions (\ref{cr16}) and (\ref{cr116}), one can rewrite the equation as:
\be
&{}&\f{1}{2}\f{\pa}{\pa t}\la v^2\ra-\f{1}{2}\f{\pa}{\pa t}B_{\r\r}=\nu \pa_\gamma\pa_\gamma\la v^2\ra+\f{1}{6\r^3}\f{\pa}{\pa \r}\left(\r^3B_{\r\r\r}\right)-\f{\nu}{\r}\f{\pa}{\pa \r}\left(\r\f{\pa B_{\r\r}}{\pa \r}\right)
\label{cr121}
\en
As we are interested in the enstrophy cascade, the first term in the R.H.S. is zero due to homogeneity and the first term in the L.H.S. is zero because of energy remains conserved in 2D turbulence in the inviscid limit (and of course, it is the high Reynolds number regime that we are interested in); it cannot be dissipated at smaller scales.
Also, as we are interested in the forward cascade which is dominated by enstrophy cascade, on the dimensional grounds in the inertial region $B_{\r\r}$ (as it may depend only on $\eta$ and $\r$) may be written as:
\be
\f{\pa}{\pa t}B_{\r\r}=A\eta\r^2
\label{cr122}
\en
where $A$ is a numerical proportionality constant.
Hence, using the relation (\ref{cr122}), the equation (\ref{cr121}) reduces to the following differential equation:
\be
\f{1}{6\r^3}\f{\pa}{\pa \r}\left(\r^3B_{\r\r\r}\right)=\f{\nu}{\r}\f{\pa}{\pa \r}\left(\r\f{\pa B_{\r\r}}{\pa \r}\right)-\f{A}{2}\eta\r^2
\label{cr123}
\en
which when solved using expression (\ref{cr115}) in the limit of infinite Reynolds number ($\nu\rightarrow 0$), one gets
\be
B_{\r\r\r}=-\f{A\eta}{2}\r^3
\label{cr124}
\en
Comparing it with the expression (\ref{neww4}) for the two-point third order velocity correlation function for the isotropic and homogeneous 2D unforced turbulence in the inertial range of the forward enstrophy cascade, we determine the value of $A$ to be -1/4.
Therefore, relation (\ref{cr122}) yields
\be
\f{\pa}{\pa t}B_{\r\r}=-\f{1}{4}\eta\r^2
\label{cr125}
\en
This, to the best of our knowledge, is yet another exact new result that has to be verified experimentally and numerically to test its validity.
\\
Suppose in the homogeneous isotropic fully-developed turbulence in two-dimensional space, energy is being supplied and the mean rate of injection of energy per unit mass is denoted by $\varepsilon$.
Let us concentrate on the inverse energy cascade.
Then technically we have to proceed as before and on doing so one would re-arrive at the differential equation (\ref{cr121}); only that now the arguments would differ.
In the larger scales viscosity is not as significant and anyway we shall be interested in the infinite Reynolds number case which would mean that the last term in the R.H.S. of equation (\ref{cr121}) would go to zero.
One obviously would also set $\f{1}{2}\f{\pa}{\pa t}\la v^2\ra=\varepsilon$ and lets assume $\f{\pa}{\pa t}B_{\r\r}\approx 0$ in the inverse cascade regime, justification of which can be sought from the fact that the ultimate result that is obtained has been experimentally and numerically verified.
So, we are left with the following differential equation:
\be
&{}&\f{1}{6\r^3}\f{\pa}{\pa \r}\left(\r^3B_{\r\r\r}\right)=\varepsilon\\
\R&{}&B_{\r\r\r}=+\f{3}{2}\varepsilon\r
\label{energy_cascade_S3}
\en
where in the last step the integration constant has been set to zero to prevent $B_{\r\r\r}$ from blowing up at $\r=0$.
This equation (\ref{energy_cascade_S3}) is the expression for two-point third order velocity correlation function of the energy cascade in inertial the range.
\subsection{Second order vorticity correlation function}
Recall that:
\be
&{}&W\equiv\la\omega_1\omega_2\ra
\label{cr125}\\
\textrm{and,}\phantom{xxx}&{}&\Omega\equiv\la(\omega_2-\omega_1)(\omega_2-\omega_1)\ra
\label{cr126}
\en
and that due to homogeneity, $\Omega$ may be expressed as:
\be
\Omega=2\la\omega^2\ra-2W
\label{cr127}
\en
In the dissipation range: $v\sim\r^2$ so, $\omega\sim\r$ and hence we may, choosing a proportionality constant $b$ (say), assume:
\be
\Omega=b\r^2
\label{cr128}
\en
Using relations (\ref{cr125}), (\ref{cr127}) and (\ref{cr128}), one gets:
\be
&{}&\la\omega_1\omega_2\ra=\la\omega^2\ra-\f{b}{2}\r^2\\
\R&{}&\la(\pa_{1\a}\omega)(\pa_{2\a}\omega)\ra=2b
\label{cr129}
\en
But we know,
\be
\eta=\nu\la(\pa_{\a}\omega)(\pa_{\a}\omega)\ra
\en
So, relation (\ref{cr129}) would yield:
\be
\eta=2\nu b
\label{cr130}
\en
where, we have put $\vec{\r}_1\approx\vec{\r}_2$ in the relation (\ref{cr129}), for, being in the dissipation range, these relations are assumed to be valid for arbitrarily small $\r$.
Combining relations (\ref{cr128}) and (\ref{cr130}), one arrives at a experimentally verifiable result for two-point second order vorticity correlation function in the dissipation range:
\be
\Omega=\f{\eta}{2\nu}\r^2
\label{cr131}
\en
\subsection{Third order mixed correlation function}
Now, we wish to find two-point third order mixed correlation function in the inertial range of enstrophy cascade.
We start by defining a two-point third order mixed correlation tensor in inertial range:
\be
&{}&\Omega_\b\equiv\la(v_{2\b}-v_{1\b})(\omega_2-\omega_1)(\omega_2-\omega_1)\ra
\label{cr132}\\
\R&{}&\Omega_\b=2M_{\b}+4W_{\b}
\label{cr133}
\en
where, $W_{\b}\equiv\la v_{1\b}\omega_1\omega_2\ra$ and $M_{\b}\equiv\la\omega_1\omega_1v_{2\b}\ra$.
Due due to isotropy and homogeneity, we can write following form for $M_{\b}$:
\be
&{}&M_{\b}=M(\r)\r_{\b}^o
\label{cr136}\\
\R&{}&\f{\pa}{\pa\r_{2\b}}M_{\b}=\la\omega_1\omega_1\pa_{2\b}v_{2\b}\ra=0
\label{cr137}\\
\R&{}&\f{\pa}{\pa\r}M({\r})+\f{M({\r})}{\r}=0\\
\R&{}& M({\r})=\f{\textrm{constant}}{\r}=0
\label{cr138}
\en
In the relation (\ref{cr137}), we are assuming incompressibility and in writing the relation (\ref{cr138}) we have taken into account the fact that $M_{\b}$ should remain finite when $\r=0$.
Relations (\ref{cr136}) and (\ref{cr138}) imply that:
\be
M_{\b}=0
\label{cr139}
\en
using which in the relation (\ref{cr133}), we get:
\be
\Omega_{\b}=4W_{\b}
\label{cr140}
\en
From the equations (\ref{cr117}) and (\ref{cr118}), we may write respectively:
\be
\f{\pa}{\pa t}\omega_{1}=-v_{1\gamma}\pa_{1\gamma}\omega_{1}+\nu\pa_{1\gamma}\pa_{1\gamma}\omega_{1}
\label{cr141}\\
\f{\pa}{\pa t}\omega_{2}=-v_{2\gamma}\pa_{2\gamma}\omega_{2}+\nu\pa_{2\gamma}\pa_{2\gamma}\omega_{2}
\label{cr142}
\en
Multiplying equations (\ref{cr141}) and (\ref{cr142}) by $\omega_2$ and $\omega_1$ respectively and adding subsequently, we get the following differential equation after averaging:
\be
\f{\pa}{\pa t}W=2\pa_{\b}W_{\b}+2\nu\pa_{\b}\pa_{\b}{W}
\label{cr143}
\en
where we have used the fact $\pa_{\b}=-\pa_{1\b}=\pa_{2\b}$.
Using relations (\ref{cr127}) and (\ref{cr140}) in the equation ({\ref{cr143}}), one gets for the inertial range for the enstrophy cascade in homogeneous, isotropic and fully-developed freely decaying turbulence in two-dimensional space in the infinite Reynolds number limit ({\it{i.e.}}, $\nu \rightarrow 0$) following differential equation:
\be
&{}&\f{\pa}{\pa t}\la\omega^2\ra-\f{1}{2}\f{\pa}{\pa t}\Omega=\f{1}{2\r}\f{\pa}{\pa\r}\left(\r\Omega_{\r}\right)
\label{cr144}\\
\R&{}&\Omega_{\r}=-2\eta\r
\label{cr145}
\en
In getting relation ({\ref{cr145}}) from the equation (\ref{cr144}), we have used the facts: $\f{1}{2}\f{\pa}{\pa t}\la\omega^2\ra=-\eta$ and $\f{1}{2}\f{\pa}{\pa t}\Omega\approx 0$ as it may be supposed that the value of $\Omega$ varies considerably with time only over an interval corresponding to the fundamental scale of turbulence and in relation to local turbulence the unperturbed flow may be regarded as steady which mean that for local turbulence one can afford to neglect $\f{\pa}{\pa t}\Omega$ in comparison with the enstrophy dissipation rate $\eta$.
 This result (relation (\ref{cr145})) has gained importance by serving as the starting point in deriving various rigorous inequalities for short-distance scaling exponents in 2D incompressible turbulence\cite{Eyink}.
\section{QG turbulence}
Quasi-geostrophic (QG) turbulence is a rather more realistic class of turbulent flow than the isotropic homogeneous 3D turbulence.
It can be seen in the large scale flows on oceans and atmosphere; thus having profound geophysical and astrophysical significance.
QG turbulence\cite{Charney} stands somewhere in between 2D and 3D turbulences.
Thus, naturally it is very appealing candidate that deserves study if one is interested in the two-dimensionalisation effect.
In the inviscid limit, besides total energy, QG flows enjoy the possession of yet another conserved quantity which is conserved at the horizontal projection of the particle motion.
We shall call this pseudo-potential vorticity to distinguish it from the potential vorticity that is conserved at a particle in a homentropic fluid.
Defining pseudo-potential enstrophy as half the square of the pseudo-potential vorticity, one would say that like 2D turbulence there are two cascades --- forward cascade of pseudo-potential vorticity and inverse cascade of energy --- in QG turbulence which, however, is inherently three dimensional in nature.
\\
Recently, a paper\cite{Lindborg} has calculated some structure functions in QG turbulence and has made illuminating revelation that isotropy in the sense of Charney\cite{Charney} is useless in deriving the structure functions for QG turbulence.
It has gone on to show that formulation of QG turbulence under the constraint of axisymmetry is productive.
However, it criticized (though somewhat rightly) the ineffectiveness of use of tensorial quantities in the case of QG turbulence in deriving the results.
Now, manipulating the tensorial quantities are at the heart of the derivation of many important two-point velocity correlation functions and other ones\cite{Landau}.
The technique is very intuitive and straightforward.
It has, recently, also been thoroughly used to find out various correlation functions for 2D turbulence\cite{SagarE0}.
In this article, we shall closely (and trickily) follow the original Kolmogorov method put forward in details in the fluid dynamics text due to Landau and Lifshitz\cite{Landau}; and repeated in the ref.-(\cite{SagarE0}), to derive structure functions in QG turbulence.
The method has the extra advantage to being able to probe into the form for the two-point third order velocity correlation function in the forward pseudo-potential enstrophy cascade regime --- this has remained uninvestigated earlier in ref.-(\cite{Lindborg}).
\subsection{Third order mixed correlation function}
First of all, we shall briefly introduce the necessary equations (see ref.-(\cite{Salmon}) for details).
Let $\vec{u}$ be the three dimensional velocity field of the fluid in a frame rotating with constant angular velocity $\vec{\Omega}$.
The fluid body (such as ocean) is assumed to be of uniform density with free surface at $z=\xi(x,y,t)$.
Suppose the bottom $z=-H(x,y)$ is rigid.
The shallow-water equations, then, are:

\be
&&\f{\pa h}{\pa t}+\vec{\nabla}.(\vec{v}h)=0
\label{swe1}\\
\textrm{and, }&&\f{D\vec{v}}{Dt}+\vec{f}\times\vec{v}=-g\vec{\nabla}\xi
\label{swe}
\en
Here, $h(x,y,t)\equiv\xi(x,y,t)+H(x,y)$, $\vec{v}\equiv(u_x,u_y)$, $\vec{\nabla}\equiv(\pa_x,\pa_y)$, $\f{D}{Dt}\equiv\f{\pa}{\pa t}+\vec{v}.\vec{\nabla}$ and $\vec{v}=\vec{v}(x,y,t)$.
$f$ is Coriolis parameter that is Taylor-expanded to write $f=f_0+\beta y$.
Using the equations (\ref{swe1}) and (\ref{swe}), one gets the relation:
\be
\f{D}{Dt}\left[\f{\hat{z}.(\textrm{{\bf curl}}\vec{u})+f}{h}\right]=0
\label{swe2}
\en
Let us assume: a) Rossby number $Ro\ll1$, b) Fractional changes in $h$ are small, and c) $\beta L/f_0\ll1$ where $L$ is the horizontal scale of the flow.
Imposing these three assumptions on the shallow-water equations one can modify the relation (\ref{swe2}) to yield
\be
\f{\pa q}{\pa t}+\vec{v}.\vec{\nabla}q=0
\label{qqqqq}
\en
where $q=\nabla^2\psi+f-f_0^2\psi/gH_0+f_0(H_0-H)/H_0$ ($\psi$ being $g\xi/f_0$) may be called pseudo-potential vorticity.
Under the same assumptions, for QG flow, one also has the condition:
\be
\vec{\nabla}.\vec{v}=0
\label{qqqq}
\en
Now, the trick is to select an arbitrary two-dimensional plane in the QG turbulent flow such that the plane's normal is along $\vec{f}$ and impose the properties of homogeneity and isotropy in the plane only.
By the way, one must keep in mind that the so-called fundamental scale of 3D turbulence has its analogy as the horizontal length scale $L$ for the case of QG turbulence. The correlation functions to be derived for the forward cascade in this article are valid in the range (which we shall call inertial range) that is much smaller than $L$ but quite larger than the scale at which the dissipation is effective. Whereas, the structure function to be derived for the inverse cascade is valid in the range whose scale is larger than the scale at which energy had been fed in. 
As we shall consider fluid bodies of uniform density only, we shall take density to be unity and let $\vec{\r}$, as usual in this article, take over the task of representing position vector in the 2D plane.
The Greek subscripts used herein can take two values $\r$ and $\p$ which respectively mean the component along the radial vector $\r$ and the component in the transverse direction.
Whenever we shall use the Latin subscript ({\it e.g.}, $a$), it should mean that it can take one more value apart from the ones mentioned above: the third value `$z$' would signify the vertical direction.
As before, Einstein's summation convention will be used extensively.
Also,
\be
\vec{\r}=\vec{\r}_2-\vec{\r}_1,\phantom{xxx}\r^o_\a\equiv\r_\a/{|\vec{\r}|},\phantom{xxx}\r^o_\r=1,\phantom{xxx}\r^o_\p=0
\label{extraqg}
\en
Now, if $\vec{v}_1$ and $\vec{v}_2$ represent the horizontal fluid velocities at the two neighbouring points at $\vec{\r}_1$ and $\vec{\r}_2$ respectively then with similar meaning for $q_1$ and $q_2$, one may define, just for the sake of notational convenience:
\be
K&\equiv&\la q_1q_2\ra
\label{25}\\
\textrm{and,}\phantom{xxx}\mega&\equiv&\la(q_2-q_1)(q_2-q_1)\ra
\label{26}
\en
The angular brackets denote an averaging procedure which averages over all possible positions of points $1$ and $2$ at a given instant of time and a given separation.
Due to homogeneity, $\mega$ may be re-expressed as:
\be
\mega=2\la q^2\ra-2K
\label{27}
\en
For simplicity, we shall take a rather idealised situation of QG turbulence which is homogeneous and isotropic on every scale in the plane.
For the unforced case, the component of the correlation tensor will obviously be dependent on time, a fact which we won't be showing explicitly in what follows.
As the features of local QG turbulence should be independent of averaged flow, the result derived below is applicable also to the local turbulence in the plane at scale $\r$ much smaller than the fundamental scale.\\
Again, we define a two-point third order mixed correlation tensor in inertial range:
\be
&{}&\mega_\b\equiv\la(v_{2\b}-v_{1\b})(q_2-q_1)(q_2-q_1)\ra
\label{32}\\
\R&{}&\mega_\b=4K_{\b}+2L_{\b}
\label{33}
\en
where just to reduce the effort of writing, we have defined:
\be
K_{\b}&\equiv&\la v_{1\b}q_1q_2\ra
\label{34}\\
\textrm{and,}\phantom{xxx}L_{\b}&\equiv&\la q_1q_1v_{2\b}\ra
\label{35}
\en
Obviously, isotropy, homogeneity and the condition (\ref{qqqq}) compels $L_{\b}$ to vanish.
Hence, equation (\ref{33}) reduces to:
\be
\mega_{\b}=4K_{\b}
\label{40}
\en
From the equation (\ref{qqqqq}), we may write for the points 1 and 2 respectively:
\be
\f{\pa}{\pa t}q_{1}=-v_{1\gamma}\pa_{1\gamma}q_{1}
\label{41}\\
\f{\pa}{\pa t}q_{2}=-v_{2\gamma}\pa_{2\gamma}q_{2}
\label{42}
\en
Multiplying equations (\ref{41}) and (\ref{42}) by $q_2$ and $q_1$ respectively and averaging subsequently after adding, we get the following differential equation:
\be
\f{\pa K}{\pa t}=2\pa_{\b}K_{\b}
\label{43}
\en
Using relations (\ref{27}) and (\ref{40}) in the equation ({\ref{43}}), one gets for the inertial range for the pseudo-potential enstrophy cascade in homogeneous and isotropic QG turbulence (forced at an intermediate scale or unforced) in inviscid limit the following differential equation:
\be
&{}&\f{\pa}{\pa t}\la q^2\ra-\f{1}{2}\f{\pa}{\pa t}\mega=\f{1}{2\r}\f{\pa}{\pa\r}\left(\r\mega_{\r}\right)
\label{44}\\
\R&{}&\mega_{\r}=-2\varepsilon_q\r
\label{45}
\en
In getting relation ({\ref{45}}) from the equation (\ref{44}), we have assumed the following:
\begin{enumerate}
\item $Q_\rho$ does not blow up at $\rho=0$.
This sets the integration constant as zero.
\item $\f{1}{2}\f{\pa}{\pa t}\la q^2\ra=-\varepsilon_q$, {\it i.e.,} there exists a pseudo-potential enstrophy sink at small scales due to some dissipative force such as viscosity and $\varepsilon_q$ is the finite and constant dissipation rate of the mean pseudo-potential enstrophy.
\item $\f{\pa}{\pa t}\mega\approx 0$ due to quasi-stationarity. It may be supposed that the value of $\mega$ varies considerably with time only over an interval corresponding to the fundamental scale of turbulence and in relation to local turbulence the unperturbed flow may be regarded as steady which mean that for local turbulence one can afford to neglect $\f{\pa}{\pa t}\mega$ in comparison with the pseudo-potential enstrophy dissipation rate $\varepsilon_q$.
\end{enumerate}
\subsection{Third order velocity correlation function}
Having explored the form for two-point third order mixed correlation function in the preceding discussion, we now proceed to find the scaling for the two-point third order velocity correlation function.
For this motive, one may define a rank two correlation tensor:
\be
B_{\a\b}\equiv\la(v_{2\a}-v_{1\a})(v_{2\b}-v_{1\b})\ra
\label{1}
\en
Isotropy and homogeneity in the plane suggests following general form for $B_{\a\b}$
\be
B_{\a\b}=A_1(\r)\delta_{\a\b}+A_2(\r)\r^o_\a\r^o_\b
\label{2}
\en
where $A_1$ and $A_2$ are functions of time and $\r$.
Making use of the relations (\ref{extraqg}) in the equation (\ref{2}), one gets:
\be
B_{\a\b}=B_{\p\p}(\delta_{\a\b}-\r^o_\a\r^o_\b)+B_{\r\r}\r^o_\a\r^o_\b
\label{3}
\en
One may expand the R.H.S. of the relation (\ref{1}) and defining $b_{\a\b}\equiv\la v_{1\a}v_{2\b}\ra$, one may proceed, keeping in mind the isotropy and the homogeneity, to arrive at:
\be
B_{\a\b}=\la v^2\ra\delta_{\a\b}-2b_{\a\b}
\label{6}
\en
Let's concentrate on the following statistically averaged quantity that will prove to be of crucial importance for deriving the desired results:
\ben
b_{\a\b,\gamma}\equiv\la v_{1\a}v_{1\b}v_{2\gamma}\ra
\enn
Invoking homogeneity and isotropy in the plane once again along with the symmetry in the first pair of indices, one may write the most general form of the third rank Cartesian tensor for this case as
\be
b_{\a\b,\gamma}&=&C(\r)\delta_{\a\b}\r^o_\gamma+D(\r)(\delta_{\gamma\b}\r^o_\a+\delta_{\a\gamma}\r^o_\b)+F(\r)\r^o_\a\r^o_\b\r^o_\gamma
\label{10}
\en
where, $C$, $D$ and $F$ are functions of $\r$.
Imposing the condition (\ref{qqqq}) on the expression (\ref{10}), one can get (in the same way as done earlier for the 2D case) the following relations:
\be
&&D=-\f{1}{2}(\r C'+C)
\label{12}\\
\textrm{and, }&&F=\r C'-C
\label{13}
\en
Here, prime ($'$) denotes derivative w.r.t. $\r$.
Defining
\be
B_{\a\b\gamma}&\equiv&\la(v_{2\a}-v_{1\a})(v_{2\b}-v_{1\b})(v_{2\gamma}-v_{1\gamma})\ra\nonumber\\
&=&2(b_{\a\b,\gamma}+b_{\gamma\b,\a}+b_{\a\gamma,\b})
\label{13-14}
\en
and putting relations (\ref{12}) and (\ref{13}) in the equation (\ref{13-14}) and using relation (\ref{10}), one gets:
\be
&{}&B_{\a\b\gamma}=-2\r C'(\delta_{\a\b}\r^o_\gamma+\delta_{\gamma\b}\r^o_\a+\delta_{\a\gamma}\r^o_\b)+6(\r C'-C)\r^o_\a\r^o_\b\r^o_\gamma
\label{14}
\en
which along with relations (\ref{10}), (\ref{12}) and (\ref{13}) yields the following expression:
\be
&{}&b_{\a\b,\gamma}=-\f{B_{\r\r\r}}{6}\delta_{\a\b}\r^o_\gamma+\f{1}{12}(\r B'_{\r\r\r}+B_{\r\r\r})(\delta_{\gamma\b}\r^o_\a+\delta_{\a\gamma}\r^o_\b)-\f{1}{6}(\r B'_{\r\r\r}-B_{\r\r\r})\r^o_\a\r^o_\b\r^o_\gamma
\label{16}
\en
The equation (\ref{swe}) suggests:
\be
\f{\pa}{\pa t}v_{1\a}=-v_{1\gamma}\pa_{1\gamma}v_{1\a}+f_{1a}\epsilon_{a\a\gamma}v_{1\gamma}-g\pa_{1\a}\xi_1
\label{17}\\
\f{\pa}{\pa t}v_{2\b}=-v_{2\gamma}\pa_{2\gamma}v_{2\b}+f_{1a}\epsilon_{a\b\gamma}v_{2\gamma}-g\pa_{2\b}\xi_2
\label{18}
\en
multiplying equations (\ref{17}) and (\ref{18}) with $v_{2\b}$ and $v_{1\a}$ respectively and adding subsequently, one gets the following:
\be
\f{\pa}{\pa t}\la v_{1\a}v_{2\b}\ra&=&-\pa_{1\gamma}\la v_{1\gamma}v_{1\a}v_{2\b}\ra-\pa_{2\gamma}\la v_{2\gamma}v_{1\a}v_{2\b}\ra\nonumber\\
&{}&+\epsilon_{a\a\gamma}\la f_{1a}v_{1\gamma}v_{2\b}\ra+\epsilon_{a\b\gamma}\la f_{2a}v_{2\gamma}v_{1\a}\ra\nonumber\\
&{}&-g\pa_{1\a}\la \xi_1 v_{2\b}\ra-g\pa_{2\b}\la \xi_2 v_{1\a}\ra
\label{19}
\en
Due to isotropy, the correlation function $\la \xi_1\vec{v}_2\ra$ should have the form $f(\r)\vec{\r}/|\vec{\r}|$.
This $f$ should not be confused with the Coriolis parameter.
But since, $\pa_{\a}\la \xi_1{v}_{2\a}\ra=0$ owing to the relation (\ref{qqqq}), $f(\r)\vec{\r}/|\vec{\r}|$ must have the form $k\vec{\r}/|\vec{\r}|^2$, where $k$ is a constant.
Now, k must vanish to keep correlation functions finite even at $\r=0$.
Thus, equation (\ref{19}) can be written as:
\be
\f{\pa}{\pa t}b_{\a\b}=\pa_{\gamma}(b_{\a\gamma,\b}+b_{\b\gamma,\a})+f_0\epsilon_{z\a\gamma}b_{\gamma\b}+f_0\epsilon_{z\b\gamma}b_{\a\gamma}
\label{20}
\en
Here we have used the approximation: $\vec{f}=f_0\hat{z}$.
Using equations (\ref{6}) and (\ref{16}), one can rewrite equation (\ref{20}) as:
\be
\f{1}{2}\f{\pa}{\pa t}\la v^2\ra-\f{1}{2}\f{\pa}{\pa t}B_{\r\r}=\f{1}{6\r^3}\f{\pa}{\pa \r}\left(\r^3B_{\r\r\r}\right)
\label{21}
\en
Note that the terms containing the Levi-Civita symbol vanish by virtue of the joint effect of the expressions (\ref{3}) and (\ref{6}), and the antisymmetry property of Levi-Civita symbol.
As we are interested in the pseudo-potential enstrophy cascade, the first term in the L.H.S. is zero because of energy remains conserved in QG turbulence in the inviscid limit: it cannot be dissipated at smaller scales.
Also, as we are interested in the forward cascade which is dominated by pseudo-potential enstrophy cascade, on the dimensional grounds in the inertial range $B_{\r\r}$ (if it is assumed to depend only on $\varepsilon_q$ and $\r$) may be written as:
\be
\f{\pa}{\pa t}B_{\r\r}=\Gamma\varepsilon_q\r^2
\label{22}
\en
where $\Gamma$ is a numerical proportionality constant.
Hence, using the relation (\ref{22}), the equation (\ref{21}) reduces to the following differential equation:
\be
\f{1}{6\r^3}\f{\pa}{\pa \r}\left(\r^3B_{\r\r\r}\right)=-\f{\Gamma}{2}\varepsilon_q\r^2
\label{23}
\en
which when solved, imposing finiteness of $B_{\r\r\r}$ for $\r=0$, gives:
\be
B_{\r\r\r}=-\f{\Gamma\varepsilon_q}{2}\r^3
\label{24}
\en
The relation (\ref{24}) is the expression for the two-point third order correlation function in the isotropic and homogeneous plane of QG turbulence (forced or unforced) in the range of the forward cascade where there is no overlapping with energy cascade.
Since $\Gamma$ has not been determined, one must confess that the equation (\ref{24}) is just a scaling law at this stage.
\\
Now suppose the fluid body is being forced at small scales {\it i.e.}, energy is being supplied and the mean rate of injection of energy per unit mass is denoted by $\varepsilon_u$ (assumed finite and constant).
Let us focus on the inverse energy cascade.
Then technically we have to proceed just as before to finally arrive at the differential equation (\ref{21}).
One obviously would set $\f{1}{2}\f{\pa}{\pa t}\la v^2\ra=\f{2}{3}\varepsilon_u$ invoking the hypothesis\cite{Charney} that there should be equipartition of energy between potential energy and the energy content in each of the two horizontal velocity components in the plane.
This equipartition had been proposed in view of the assumption that at sufficiently small scales the interaction of the mean flow with the eddies (and thus the eddy-energies) diminishes; as a result, for increasingly smaller vertical and horizontal scales the energies will tend to become homogeneous and equally distributed among the perturbations.
By the way, the concept of equipartition of energy is very old and wide-spread in the literature of statistical mechanics.
Historically, equilibrium statistical mechanics had been used to justify many aspects of turbulence {\it e.g.,} the dual cascades in 2D turbulence {\it etc.}
A detailed discussion may be found in the books by Chorin\cite{Chorin} and Lim {\it et al.}\cite{Lim}.
Now, lets also assume that $\f{\pa}{\pa t}B_{\r\r}\approx 0$ in the inverse cascade regime supposing the forced QG turbulence to be in the state of quasi-stationarity.
So we are left with the following differential equation:
\be
&{}&\f{1}{6\r^3}\f{\pa}{\pa \r}\left(\r^3B_{\r\r\r}\right)=\f{2}{3}\varepsilon_u
\label{derteqg}\\
\R&{}&B_{\r\r\r}=+\varepsilon_u\r
\label{energy_cascade_S3qg}
\en
where in the last step the integration constant has been set to zero to prevent $B_{\r\r\r}$ from blowing up at $\r=0$.
The expression (\ref{energy_cascade_S3qg}) is the expression for the two-point third order velocity correlation function in the isotropic and homogeneous plane of forced QG turbulence for the inverse energy cascade.
\\
The fact that the structure functions for the inherently three-dimensional QG turbulence are more like that for the 2D turbulence than that for the 3D turbulence speaks volumes for the importance of study of third order structure functions for demystifying the two-dimensionalisation effect of the 3D turbulent fluid due to rapid rotation.
This serves as the motivation for jumping into the subject of rotating flows and to attempt finding the form of $S_3$ therein.
\section{Rotating turbulence}
All the studies on the two-dimensionalisation effect are mainly for low $Ro$ high $Re$ limit while the high $Ro$ and high $Re$ limit has been rather less ventured in relation to the two-dimensionalisation effect of turbulence, although the second case, we believe, should be analytically more tractable.
If, using calculations of structure functions, in the limit of high $Ro$ and high $Re$, one wishes to see whether a trend towards two-dimensionalisation of 3D homogeneous isotropic turbulence occurs or not, then basically one would have to check (a) if $S_3=-(4/5)\varepsilon l$ at small scales for 3D turbulence shows a tilt towards $S_3=(3/2)\varepsilon l$ at large scales for the 2D turbulence and (b) if the forward energy cascade is depleted at the smaller scales.
As we shall show, in the lowest order calculation this is what one may get, hinting at the initiation of the effect of two-dimensionalisation of 3D turbulence owing to the small anisotropy induced by slow rotation.
\subsection{Relevant scales}
Let us look in to the various length scales that have to be taken into consideration while talking about a homogeneous rotating turbulence which basically satisfies following version of Navier-Stoke's equation:
\be
\f{\pa\vec{v}}{\pa t}+\left(\vec{v}.\vec{\na}\right)\vec{v}&=&-\f{1}{\rho}\vec{\na}P-\vec{\Omega}\times\left(\vec{\Omega}\times\vec{x}\right)-2\vec{\Omega}\times\vec{v}+\nu\na^2\vec{v}+\vec{f}
\label{nse}
\en
In this context $\vec{f}$ is external force and $\vec{\Omega}$ is angular velocity.
Various parameters to be considered are: $\nu$ (kinematic viscosity), $\varepsilon$ (finite mean rate of dissipation of energy per unit mass), $\Omega$ (angular velocity) and $l_0$ (integral scale which typically is the system-size).
The three important time-scales involved in the system are: $t_l\sim\varepsilon^{-1/3}l^{2/3}$ (eddy-turnover time or circulation time for the eddy of scale $l$; $l\le l_0$), $t_{\Omega}\sim\Omega^{-1}$ and $t_d\sim l^2/\nu$ (diffusion time scale).
It is well-known that a length scale $l_\Omega=\sqrt{(\varepsilon/\Omega^3)}$ is what responsible for the estimation of the anisotropy introduced by the rotation.
The competition between the time-scales $t_l$ and $t_d$ gives rise to what is known as dissipation length scale $l_d$, defined as $l_d=(\nu^3/\varepsilon)^{1/4}$ and a similar competition between the time-scales $t_d$ and $t_\O$ allows us to define a length scale $l_{\O d}=\sqrt{(\nu/\O)}$.
Now, lets look at the typical scenario when $Ro$ is moderate.
The four vital length scales are typically arranged according to the order : $l_0>l_\O>l_{\O d}>l_d$.
Thus, the regime $l_0>l>l_\O$ is the regime where effect of rotation is important and anisotropy reigns.
The scales $l\in(l_\O,l_d)$ may be considered to have isotropy, though to be precise, probably $l_d$ here should be replaced by $l_{\O d}$ since rotation seems to be bringing the effect of viscosity to rather larger length scales.
So, now what happens when the $Ro$ is decreased by increasing the angular velocity is interesting.
Both the scales $l_\O$ and $l_{\O d}$ rush towards the dissipation length scale, thereby increasing the anisotropic regime and at the angular velocity $\O=\O_a\equiv\sqrt{(\varepsilon/\nu)}$ one has $l_\O=l_{\O d}=l_d$ and the turbulence is fully anisotropic.
\\
Strictly speaking, even a small rotation introduces anisotropy (however small) at all scales and the isotropic regime does have a degree of anisotropy in it as we shall see shortly.
In the fully anisotropic limit, {\it i.e.} for $\O=\O_a$, one expects full decoupling of the plane perpendicular to the rotation axis from the direction of the rotation axis.
However, even in the partially anisotropic limit ({\it e.g.} when we have slow rotation imparted on the turbulent fluid), $l_z$ should still be given a special status for being in the direction of the rotation axis, by which we mean that the structure functions should no longer depend on $l$ but rather on $l_z$ and $\vec{l}_\p$ (where $l^2=l_z^2+l_\p^2$ and $|\vec{\O}|=\O_z$).
\\
We shall see how this decoupling sets in, in the limit of low angular velocity and try to study in that very limit, the two-point third order structure function in the first approximation and see how the effect of two-dimensionalisation is all set to sneak in with the switching on of rotation.
\subsection{$S_3$ for small $\O$}
Let us start with low $\O$-limit.
With this statement we mean, as discussed in the previous subsection, $\O\ll\O_a$. 
So, the entire fluid may still be treated as isotropic but as rotation should play a role, we assume that $\la v_iv_jv'_k\ra$ (where angular brackets mean ensemble average and $v_i=v_i(\vec{x},t)$ is the $i$-th component of velocity and similarly, $v'_i=v_i(\vec{x}+\vec{l},t)$) should depend on $\vec{\O}$ as well.
$\vec{\O}$ would take care of the mild anisotropy.
Since, physically speaking, $S_3$ should not depend on which way the rotation axis is and since we are interested in low values of $\O$, we shall let $\la v_iv_jv'_k\ra$ depend only on the terms quadratic in $\O$ and not bother about higher order terms in $\O$.
As a result, we write the following most general tensorial form for $\la v_iv_jv'_k\ra$:
\be
b_{ij,k}&\equiv&\la v_iv_jv'_k\ra\\
&=&C(l)\delta_{ij}l^o_k+D(l)(\delta_{ik}l^o_j+\delta_{jk}l^o_i)+F(l)l^o_il^o_jl^o_k\nonumber\\
&&+G(l)[(\epsilon_{imk}l^o_j+\epsilon_{jmk}l^o_i)l^o_m]+H(l)\O_i\O_jl^o_k\nonumber\\
&&+I(l)[(\epsilon_{imk}\O_j+\epsilon_{jmk}\O_i)\O_m]+K(l)(\O_i\O_kl^o_j+\O_j\O_kl^o_i)
\label{rev2}
\en
where $l_i^o$ is the $i$-th component of the unit vector along $\vec{l}$. We have assumed that the coefficients are dependent only on $l$ and it is the $\vec{\O}$ which is taking care of the mild anisotropy which the turbulent fluid might have.
We must accept that the assumption of letting coefficients depend only on $l$ is rather crude in the light of the complex forms that the two-point tensors in a fully anisotropic turbulence flow take\cite{Biferalee}.
The justification, and hence solace, for the assumption can be drawn from the fact that very simple revealing results matching with recent experiments are arrived at in the long run.
As we are considering incompressible fluid, we must have:
\be
\pa'_kb_{ij,k}=0
\label{rev3}
\en
which when applied to relation (\ref{rev2}), yields relationships between various coefficients.
As usual Einstein summation convention has been extensively followed in these calculations unless otherwise specified.
Using relations (\ref{rev2}) and (\ref{rev3}), one lands up in the end on the following:
\be
B_{ijk}&\equiv&\la(v'_i-v_i)(v'_j-v_j)(v'_k-v_k)\ra\\
&=&2(b_{ij,k}+b_{jk,i}+b_{ki,j})\\
&=&-2(lC'+C)(\delta_{ij}l^o_k+\delta_{ik}l^o_j+\delta_{jk}l^o_i)+6(lC'-C)l^o_il^o_jl^o_k\nonumber\\
&&+4Jl(\O_i\O_jl^o_k+\O_i\O_kl^o_j+\O_j\O_kl^o_i)
\label{rev4}
\en
Here, in expression (\ref{rev4}), prime (``$'$'') denotes derivative w.r.t. $l$ and $J$ is a constant which, curiously enough, is of the same dimension $[L^2T^{-1}]$ as that of the kinematic viscosity.
Now we can see that using the relation (\ref{rev4}), two-point third order structure function ($S_3$) can be extracted from $B_{ijk}$ in the following way:
\be
S_3(l)&\equiv&\la(\delta v_{\parallel}(\vec{l}))^3\ra\equiv\left\la\left[\left\{\vec{v}(\vec{x}+\vec{l})-\vec{v}(\vec{x})\right\}.\f{\vec{l}}{l}\right]^3\right\ra\\
\R S_3(l)&=&\la([v'_i-v_i)l^o_i][(v'_j-v_j)l^o_j][(v'_k-v_k)l^o_k]\ra\\
\R S_3(l)&=&B_{ijk}l^o_il^o_jl^o_k\\
\R S_3(l)&=&-12C+\f{12J}{l}(\vec{\O}.\vec{l})^2
\label{rev5}
\en
where we have used relation (\ref{rev4}).
One may define physical space energy flux ($\varepsilon(\vec{l})$) as:
\be
&&\varepsilon(l)\equiv-\f{1}{4}\vec{\nabla}_l.\la |\delta\vec{v}(\vec{l})|^2\delta\vec{v}(\vec{l})\ra
\label{star}\\
\Rightarrow&&\varepsilon(l)=lC''+7C'+\f{8C}{l}+3J\O^2+\f{6J}{l^2}(\vec{\O}.\vec{l})^2
\label{rev6}
\en
To get relation (\ref{rev6}), we have again made use of the relation (\ref{rev4}).
The energy flux through the wave number $K$ ($\Pi_K$) for the isotropic homogeneous turbulence may be calculated to be:
\be
\Pi_K=\f{2}{\pi}\int_0^\infty dl\f{\sin(Kl)}{l}(1+l\pa_l)\varepsilon(l)
\label{rev7}
\en
Now if one makes the standard assumption (often made during the derivation of $S_3$) that as $Re\rightarrow\infty$, the mean energy dissipation per unit mass $\varepsilon(\nu)$ tends to a positive finite value ({\it i.e.,} $\lim_{\nu\rightarrow 0}\varepsilon(\nu)=\varepsilon>0$), then $\lim_{\nu\rightarrow 0}\Pi_K=\varepsilon$ in the inertial regime.
Therefore, in the inertial range, putting $x=Kl$, one has
\be
\Pi_K=\f{2}{\pi}\int_0^\infty dx\f{\sin(x)}{x}f\left(\f{x}{K}\right)=\varepsilon
\label{rev8}
\en
where,
\be
f\left(\f{x}{K}\right)=f(l)=(1+l\pa_l)\varepsilon(l)
\label{rev9}
\en
For small $l$ (large $K$), the integral in relation (\ref{rev8}) yields
\be
f(l)\approx\varepsilon
\label{rev10}
\en
Now using relations (\ref{rev6}), (\ref{rev9}) and (\ref{rev10}), we form a differential equation which when solved, keeping in mind that $S_3$ should not blow up at $l=0$, one gets following form for $S_3$ in slowly rotating homogeneous turbulent fluid. 
\be
S_3(l)=-\f{4}{5}\varepsilon l+\f{12}{5}Jl[\O^2+7(\O_kl_k^o)^2]
\label{rev11}
\en
One may note from the relation (\ref{rev11}) that how magically $\O$ has brought up the anisotropic effects even for small $\O$ though for the entire calculation we followed the procedure meant for the homogeneous isotropic turbulence.
Thus, the form for $S_3$ is plausible.
\\
One may ask: Does the effect of two-dimensionalisation shows up in the relation (\ref{rev11})?
As one may note from the relation (\ref{rev11}) this is quite a possibility but the only catch being that $J$ should be positive --- an issue which we have not been able to resolve.
If $J$ is positive, it means if we increase $\O$ the value of $S_3$ would distort away from the usual $-(4/5)\varepsilon l$ for the non-rotating case to more positive values.
This apparently shows that the effective value of $\varepsilon$ is decreased depicting that the forward energy transfer is depleted which is in keeping with what is expected and hence the tendency of the rotating 3D turbulence to show the effect the two-dimensionalisation is being highlighted.
That the sign of $J$ should be positive is a question remains to be addressed.
\\
By the way, the relation (\ref{rev11}) also suggests that the coefficients in the tensorial form for $b_{ij,k}$ should have dependence on $l_z$ and $l_\p$ separately effecting a mild decoupling of directions.
So taking hint from it, we proceed to rewrite $b_{ij,k}$ for slowly rotating 3D turbulent fluid but now introducing anisotropy directly into the coefficients and not letting $\O$ take care of anisotropy explicitly.
Of course, the coefficients will now depend on $\O$.
\\
For completely isotropic homogeneous turbulence, one would write following general form (relation (\ref{rev1000})) for $b_{ij,k}$ which is made up of Kronecker delta and components of the unit vectors $\vec{l}/|\vec{l}|$.
\be
b_{ij,k}&=&C(l)\delta_{ij}l^o_k+D(l)(\delta_{ik}l^o_j+\delta_{jk}l^o_i)+F(l)l^o_il^o_jl^o_k
\label{rev1000}
\en
The expression is symmetric in $i$ and $j$ and the coefficients are dependent on $l$ only.
As discussed earlier, with rotation coming into effect, anisotropy comes into effect.
If this effects in the possible decoupling (even if partial) of the direction along the rotation axis (which we shall take along the z-axis), then mathematically we may introduce this effect by modifying the form (\ref{rev1000}) of $b_{ij,k}$ to the following:
\be
b_{ij,k}&=&C(l,l_z,\O)\delta_{ij}l^o_k+D(l,l_z,\O)(\delta_{ik}l^o_j+\delta_{jk}l^o_i)+F(l,l_z,\O)l^o_il^o_jl^o_k
\label{rev12}
\en
If one uses the incompressibility condition (relation (\ref{rev3})), one gets:
\be
&&D=\f{l}{2}(-C'-\f{\dot{C}l_z}{l})-C
\label{rev50}\\
\textrm{and}\phantom{xxx}&&\dot{D}=0
\label{rev60}
\en
where dot represents the derivative w.r.t. $l_z$ and prime, as before, the derivative w.r.t. $l$.
Using equation (\ref{rev50}) in the equation (\ref{rev60}), one land up on:
\be
&&\ddot{C}l_z+l\dot{C}'+3\dot{C}=0\\
&\Rightarrow&C=\sum_n A_nl^{-n-2}l_z^n\\
&\Rightarrow&C\ne 0\phantom{x}\textrm{for}\phantom{x}n\in(-\infty,-2]\cap[0,\infty)\\
&\Rightarrow&C=D=F=0
\label{rev70}
\en
In arriving at the result (\ref{rev70}), we have taken care of the fact that $C$ can not be allowed to blow up for either for $l_z=0$ or for $l=0$.
Thus, relation (\ref{rev12}) vanishes trivially.
So, we are left with the following choice:
\be
b_{ij,k}&=&C(l_\p,l_z,\O)\delta_{ij}l^o_k+D(l_\p,l_z,\O)(\delta_{ik}l^o_j+\delta_{jk}l^o_i)+F(l_\p,l_z,\O)l^o_il^o_jl^o_k
\label{rev13}
\en
Using equations (\ref{rev3}) and (\ref{rev13}), we arrive at following relationship between the coefficients:
\be
D&=&-\f{l_\p}{2}\td{C}-\f{l_z}{2}\dot{C}-C\\
F&=&\f{l^2}{2}\td{\td{C}}+\f{l^2l_z}{2l_\p}\dot{\td{C}}+\left(\f{3l^2}{2l_\p}-\f{l_\p}{2}\right)\td{C}-\f{l_z}{2}\dot{C}-C
\label{rev14}
\en
Here tilde and dot define derivatives w.r.t. $l_\p$ and $l_z$ respectively.
Proceeding monotonously as before we get
\be
B_{ijk}&=&2(b_{ij,k}+b_{jk,i}+b_{ki,j})\\
&=&-2(l_\p\td{C}+l_z\dot{C}+C)(\delta_{ij}l^o_k+\delta_{ik}l^o_j+\delta_{jk}l^o_i)+6Fl^o_il^o_jl^o_k
\label{rev15}
\en
And hence,
\be
S_3=B_{ijk}l^o_il^o_jl^o_k=6[F-(l_\p\td{C}+l_z\dot{C}+C)]
\label{rev16}
\en
The definition for the physical space energy flux ($\varepsilon(\vec{l})$) has to be obviously modified.
Natural choice would be:
\be
\la |\delta\vec{v}(\vec{l})|^2\delta\vec{v}(\vec{l})\ra
=B_{ii\alpha}l^o_\alpha\f{\vec{l}_\p}{l_\p}+B_{iiz}l^o_z\f{\vec{l}_z}{l_z}
\label{rev17}
\en
where $\alpha$ takes two values: $x$ and $y$ only.
Now, using relations (\ref{star}), (\ref{rev14}), (\ref{rev15}) and (\ref{rev17}) and performing tedious algebra one gets:
\be
\varepsilon(l_\p,l_z)&=&\f{-1}{4(l_\p^2+l_z^2)^2}\left[(3l_\p^6+6l_\p^4l_z^2+3l_\p^2l_z^4)\td{\td{\td{C}}}\right.\nonumber\\
&&+\left(3l_\p^5l_z+6l_\p^3l_z^3+3l_\p^4l_z^2+6l_\p^2l_z^4+3l_\p l_z^5+3l_z^6\right)\dot{\td{\td{C}}}\nonumber\\
&&+\left(3l_\p^3l_z^3+6l_\p l_z^5+3l_\p^{-1} l_z^7\right)\ddot{\td{C}}\nonumber\\
&&+\left(5l_\p^5+6l_\p^4 l_z+23l_\p^3 l_z^2+12l_\p^2 l_z^3+18l_\p l_z^4+6l_z^5\right)\td{{\td{C}}}\nonumber\\
&&+\left(-7l_\p^4 l_z+5l_\p^3 l_z^2-l_\p^2 l_z^3+23l_\p l_z^4+6l_z^5+18l_\p^{-1} l_z^6\right)\dot{\td{C}}\nonumber\\
&&+\left(-12l_z^4-8l_\p^3 l_z-20l_\p^2 l_z^2+36l_\p l_z^3+18l_z^4+8l_\p^{-1} l_z^5\right)\td{C}\nonumber\\
&&+\left(-13l_\p^3 l_z-43l_\p^2 l_z^2-39l_\p l_z^3-17l_z^4\right)\dot{C}\nonumber\\
&&+\left.\left(-4l_\p^3-8l_\p^2l_z-12l_\p l_z^2\right)C\right]
\label{rev18}
\en
The energy flux ($\Pi_K$) through the wave number $K$ for the homogeneous (not necessarily isotropic) turbulence may be shown to be:
\be
\Pi_K=\f{1}{2\pi^2}\int_{\m{R}^3}d^3l\f{\sin(Kl)}{l}\vec{\nabla}_l.\left[\varepsilon(\vec{l})\f{\vec{l}}{l^2}\right]
\label{rev19}
\en
Using cylindrical polar coordinates we reduce the relation (\ref{rev19}) to:
\be
\Pi_K&=&\f{1}{\pi}\int\int l_\p dl_\p dl_z\left\{\f{\sin(Kl)}{l}\left[\f{l_\p}{l^2}\f{\pa}{\pa l_\p}+\f{l_z}{l^2}\f{\pa}{\pa l_z}+\f{1}{l^2}\right]\varepsilon(\vec{l})\right\}
\label{rev20}
\en
Now, we introduce the variables $y=Kl_\p$ and $z=Kl_z$ in relation (\ref{rev20}) to get:
\be
\Pi_K=\f{1}{\pi}\int_{z=-\infty}^{\infty}\int_{y=0}^\infty dydz\f{\sin(y^2+z^2)^{\f{1}{2}}}{y^2+z^2}\left[f\left(\f{y}{K},\f{z}{K}\right)\right]\nonumber\\
\label{rev21}
\en
Now, let's probe small $l$ behaviour.
Because $\int_{z=-\infty}^{\infty}\int_{y=0}^\infty dydz[\sin(y^2+z^2)^{1/2}]/(y^2+z^2)=\pi^2/2$, we have
\be
f(l_\p,l_z)\approx \f{2\varepsilon}{\pi}
\label{rev22}
\en
Obviously, $\varepsilon$ has the meaning of finite positive mean rate of dissipation of energy per unit mass. 
Using the expressions (\ref{rev18}) and (\ref{rev22}), we look for the $l_z=0$ limit.
One then has the result:
\be
&&\left[l_\p\f{\pa}{\pa l_\p}+1\right](3l_\p^2\td{\td{\td{C}}}+5l_\p\td{\td{C}}-12\td{C}-4\f{C}{l_\p})=-\f{8\varepsilon}{\pi}\\
\Rightarrow&&3l_\p^4\td{\td{\td{\td{C}}}}+14l_\p^3\td{\td{\td{C}}}-2l_\p^2\td{\td{C}}-16{l_\p}\td{C}=-\f{8\varepsilon}{\pi}l_\p\\
\Rightarrow&&C=\left(A_1+A_2l_\p^{-1}+A_3l_\p^{\f{7-\sqrt{97}}{6}}+A_4l_\p^{\f{7+\sqrt{97}}{6}}\right)+\f{\varepsilon l_\p}{2\pi}
\label{rev100}
\en
Relations (\ref{rev14}), (\ref{rev16}) and (\ref{rev100}) together yield following expression for $S_3$:
\be
&&S_3|_{l_z=0}=-\f{6}{\pi}\varepsilon l_\p+A_4\left[3\left(\f{7+\sqrt{97}}{6}\right)\left(\f{1+\sqrt{97}}{6}\right)-12\right]l_\p^{\f{7+\sqrt{97}}{6}}\\
\Rightarrow&&S_3|_{l_z=0}=-\f{6}{\pi}\varepsilon l_\p+Al_\p^{\f{7+\sqrt{97}}{6}}
\label{rev23}
\en
where, $A$ is a constant which for obvious reason depends on $\O$ and $\varepsilon$.
Using dimensional arguments and introducing a non-dimensional constant $c$, we may set
\be
A=c\O^{\f{1+\sqrt{97}}{4}}\varepsilon^{\f{11-\sqrt{97}}{12}}
\label{rev24}
\en
From relations (\ref{rev23}) and (\ref{rev24}), we may write finally
\be
S_3|_{l_z=0}=-\f{6}{\pi}\varepsilon l_\p+c\O^{\f{1+\sqrt{97}}{4}}\varepsilon^{\f{11-\sqrt{97}}{12}}l_\p^{\f{7+\sqrt{97}}{6}}
\label{rev25}
\en
This (relation (\ref{rev25})) is the final form for two-point third order structure function in the plane whose normal is parallel to the rotation axis for slowly rotating homogeneous 3D turbulence.
\subsection{Energy spectrum for small $\O$}
If we for the time being forget about the issue of anomalous scaling, then a mere inspection of the relation (\ref{rev25}) from the point of view of dimensional analysis would tell that in the directions perpendicular to the axis of rotation, there are two possible energy spectrums {\it viz.}
\be
&&E(k)\sim k^{-\f{5}{3}}
\label{rev26}\\
\textrm{and}, \phantom{xxx}&&E(k)\sim k^{-\f{16+\sqrt{97}}{9}}
\label{rev27}
\en
which are respectively due to the first term and the second term in the R.H.S. of the relation (\ref{rev25}).
It is very interesting to note that the exponent of $k$ in the relation (\ref{rev27}), {\it i.e.} $-(16+\sqrt{97})/9$, equals $-2.87$ which is in between $-3$ (for 2D turbulence) and $-2$ (for rapidly rotating 3D turbulence as proposed by Zhou).
Obviously, the spectrum (\ref{rev26}) will be dominant compared to the spectrum (\ref{rev27}).
But as the $\O$ is increased (of course, remaining within a range so that the anisotropy is not strong enough to breakdown the arguments used to calculate the $S_3$ of the relation (\ref{rev25})), the spectrum (\ref{rev27}) becomes more and more prominent; thereby two-dimensionalisation of the 3D homogeneous turbulent fluid is initiated which then carries over to high rotation regime as is being extensively studied now-a-days.
This signature of two-dimensionalisation is, of course, in agreement with what present literature on turbulence hails as the two-dimensionalisation of turbulence.
Thus, the third order structure function has proved to be very handy in studying this effect because the phenomenon of two-dimensionalisation is reflected as a change in the scaling law of the third order structure function.
We pause here for a moment and ponder upon the signatures of two-dimensionalisation effect in a rather more intuitive, though a bit non-rigorous, way.
\subsection{The signatures: an intuitive picture}
Let us first concentrate on why at all there should be an inverse cascade of energy.
Inverse cascade of energy is a trademark of 2D turbulence where a second conserved quantity --- enstrophy --- besides energy plays the defining role behind it.
One might be tempted to search for this conserved quantity in the case of rapidly rotating 3D turbulence, for, there in the limit of infinite rotation the axes of all the vortices are expected to point up towards the direction of angular velocity.
Hence, looking at the every section perpendicular to the axis one might tend to think that 2D turbulence is being shown by each transverse section.
This obviously is not a correct inference because of the non-zero axial velocity may depend on the coordinates on the plane.
Searching for the enstrophy conservation seems to be a dead end as far as explaining the inverse cascade in rapidly rotating turbulence is concerned.
In such an unfortunate scenario, helicity (defined as $\int{\vec{v}}.{\vec{\omega}}d^3{\vec{r}}$) which remains conserved in a 3D inviscid unforced flow comes to our rescue.
It has been long known that helicity is introduced into a rotating turbulent flow\cite{Brissaud}.
Kraichnan\cite{Kraichnan} argued that both the helicity and energy cascade in 3D turbulence would proceed from lower to higher wave numbers and went on to remark that forward helicity cascade would pose a hindrance for the energy cascade --- a fact validated by numerical simulations\cite{Andre,Polifke}.
He also showed that in presence of helicity two-way cascade is possible.
Lets see topologically why this should be so.
It is well known that a knotted vortex tube is capable of introducing helicity in fluid\cite{Moffatt}.
Consider a knotted vortex tube (see Fig-1) in a turbulent flow.
%
%%%%%%%%%%%%%%%%%%%%%%%%%%%%%%%%%%%%%%%%%%%%%%%%%%%%%%%%%%%%%%%%%%%%%%%%%%%%%%%%%%%%%%%%%%%%%%%%%%%%%%%%%%%%%%%%%%%%%%%%%%%%
\begin{figure}
\centering
\includegraphics[width=7.0cm]{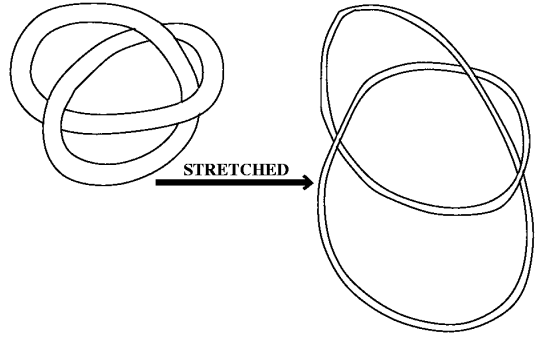}
\caption{A knotted vortex tube. When it is stretched the tube thins out to create smaller eddies but the entire structure occupies a larger volume.}
\end{figure}
%%%%%%%%%%%%%%%%%%%%%%%%%%%%%%%%%%%%%%%%%%%%%%%%%%%%%%%%%%%%%%%%%%%%%%%%%%%%%%%%%%%%%%%%%%%%%%%%%%%%%%%%%%%%%%%%%%%%%%%%%%%%
%
Due to the vortex stretching phenomenon in turbulence, the vortex line stretches and as a result owing to the assumed incompressibility of the fluid the tube thins out keeping the volume inside it preserved and smaller scales are created; in a sense, this is what is meant by the flow of energy to the smaller scales.
But now this also means that the ``scale'' of the knotted structure would in general increase {\it i.e.,} the knot would now reach out to farther regions in the fluid.
Evidently, if we wanted this scale to reduce, we must let the stretched knotted tube fold in such a way so that the scale becomes smaller; such a neat arrangement seems to be a far cry in a turbulent flow which is inherently chaotically random causing the separation of two nearby particles of fluid on an average.
Thus, as the degree of knottedness measures helicity, the aforementioned argument suggests that if one forces energy to go to smaller scales, helicity would tend to go to larger scale and vice-versa.
This topological argument gives an intuitive way of comprehending how the forward helicity cascade can inhibit the forward cascade of energy.
The point is that in presence of forward helicity cascade, reverse cascade of energy is not impossible.
\\
Waleffe\cite{waleffe2}, with the help of detailed helicity conservation by each triad, showed that helicity indeed affects the turbulence dynamics even in isotropic turbulence; this is a kind of catalytic effect.
One can thus take inspiration to make the argument in the previous paragraph more concrete by playing around with a simplified triad using logic in the line suggested by Fjortoft's theorem\cite{Fjortoft} in 2D turbulence.
Let the helicity spectrum be $H(k)$ and the energy spectrum be $E(k)$.
It may be shown that 
\be
|H(k)|\le kE(k)
\label{condition}
\en
Consider 3D Euler equation in Fourier space truncated in order to retain only three parallel wave vectors $\vec{k}_1$, $\vec{k}_2$ and $\vec{k}_3$ and suppose it is possible for these three particular wave vectors to be such that $|H(k)|=nkE(k)$, where $n$ is a positive number lesser than 1 to be in consistence with the relation (\ref{condition}).
Assume $\vec{k}_2=2\vec{k}_1$ and $\vec{k}_3=3\vec{k}_1$.
Conservation of energy and helicity imply that between two times $t_1$ and $t_2$, the variation $\delta E_i=E(k_i,t_2)-E(k_i,t_1)$ satisfies two constraints
\be
\delta E_1+\delta E_2+\delta E_3=0
\label{energy1}\\
\textrm{and, }nk_1\delta E_1+nk_2\delta E_2+nk_3\delta E_3=0
\label{helicity1}
\en
solving which in terms of $\delta E_2$, we get:
\be
\delta E_1=\delta E_3=-\f{\delta E_2}{2}
\label{energy2}\\
\textrm{and, }nk_1\delta E_1=-\f{n}{4}k_2\delta E_2;\phantom{xxx}nk_3\delta E_3=-\f{3n}{4}k_2\delta E_2
\label{helicity2}
\en
If one assumes that the wave vector $k_2$ is losing energy, {\it i.e. $\delta E_2<0$}, then the results (\ref{energy2}) and (\ref{helicity2}) show that as more helicity goes into the higher wavenumber, the energy is equally transferred to both the lower and the higher wave numbers suggesting a possibility of the coexistence of reverse and forward energy cascades. 
\\
Now let us come to the point.
In the case of 3D isotropic and homogeneous turbulence rotation can input helicity in it when there is a mean flow in the inertial frame and this value of input helicity increases with the increase in angular velocity.
Experiments on rotating turbulence invariably introduce helicity.
As the angular velocity is increased the helicity increases enough to inhibit the energy cascade appreciably so that a reverse cascade is seen.
This consistently explains the reason behind the existence of the reverse energy cascade in a rapidly rotating turbulent flow.
Hence, the argued existence of a direct helicity cascade in such experiments turns out to be an interesting (however not rigourously proven) assumption.
\\
As discussed earlier, the next important signature of the two-dimensionalisation of turbulence that remains to be pondered upon is the exponent of the wave vector in the energy spectrum relation.
To be precise, if one wishes angular velocity becomes a relevant parameter in the energy spectrum $E(k)$, simple dimensional analysis would give:
\be
E(k)\propto\O^{\frac{3m-5}{2}}\varepsilon^{\f{3-m}{2}}k^{-m}
\label{revepl0}
\en
where $m$ is a real number.
$m$ must be restricted within the range 5/3 to 3 to keep the exponents of $\O$ and $\varepsilon$ in relation (\ref{revepl0}) non-negative.
The two limits $m=5/3$ and $m=3$ corresponds to isotropic homogeneous 3D turbulence and 2D turbulence respectively.
The spectrum due to Zhou --- $E(k)\sim k^{-2}$ --- is due to an intermediate value of $m=2$.
So, as far as the present state of the literature on rotating turbulence is concerned, two-dimensionalisation of 3D turbulence would mean the dominance of a spectrum which goes towards $E(k)\sim k^{-3}$ and which may choose to settle at $E(k)\sim k^{-2}$.
\\
Lets give a twist to the tale.
In general, the energy spectrum\cite{Brissaud} in the inertial range will be determined by both the helicity cascade and the energy cascade which simply means that the energy spectrum from the dimensional arguments should be written as
\be
E(k)\propto\varepsilon^{\f{7}{3}-m}h^{m-\f{5}{3}}k^{-m}
\label{spectrum}
\en
where $h$ is the rate of helicity dissipation per unit mass.
Demanding positivity of the exponents of $\varepsilon$ and $h$, one fixes the possible values for $k$ within the closed range [5/3,7/3], imposing which on the arguments given in the previous paragraph, one can easily propound the range
\be
2\le m\le\f{7}{3}
\label{range}
\en
for the rapidly rotating 3D turbulent flow.
Direct experiments\cite{Morize1} by Morize {\it et al.} have found energy spectrum for rapidly rotating turbulence going as $k^{-2.2}$ which is as predicted by the relation (\ref{range}).
\\
One may note that the scaling exponent derived as expression (\ref{rev27}) has not fallen into the more strict range $[-7/3,-5/3]$ obviously because $\O$ is too low and may be because to maintain isotropy to a certain extent for the sake of hiccup-free calculations we have chosen not to include terms involving $\epsilon_{ijk}$ in the relation (\ref{rev13}) which could grab the effect of helicity explicitly; thereby again showcasing the need for the helicity to be effective to give the right exponent for the rotating turbulence.
\subsection{Yet another signature}
 Having explained the two signatures of the two-dimensionalisation effect, we search for another possible signature of the effect.
The advection of a passive scalar $\theta$ may serve the purpose since the Yaglom's law\cite{Monin} in d-D incompressible turbulent fluid may be written as $\la\delta v_{\parallel}(\delta\theta)^2\ra=-({4}/{d})\varepsilon_{\theta}l$, where $\varepsilon_{\theta}\equiv\kappa\left\la\left({\pa_{l_i}}\theta\right)\left({\pa_{l_i}}\theta\right)\right\ra=-{\pa_t}\la\theta^2\ra$ and $\kappa$ being the diffusivity.
This law distinguishes between a 2D and a 3D turbulence and hence it is worth getting a form for it for a rotating 3D turbulence and find if in a plane perpendicular to the rotation axis it reduces to the form for 2D turbulence and thereby bringing in the effect of two-dimensionalisation.
Since we have witnessed earlier that small $\O$ could bring in anisotropy in the otherwise isotropic scales, one would look out for the effect of small $\O$ on the passive scalar which follows the equation:
\be
\f{\pa\theta}{\pa t}+\vec{\na}.(\vec{v}\theta)=\kappa\na^2\theta-\epsilon_{ijk}\O_j\f{\pa}{\pa x_i}(x_k\theta)
\label{ps}
\en
If one goes by the procedure given in the reference \cite{Biskamp} to find out a value for $\la\delta v_{\parallel}(\delta\theta)^2\ra$ for small $l$ in this case assuming very small $\O$ (and hence isotropy), one arrives back at the Yaglom's law.
We can however land up on a very neat experimentally and numerically verifiable correlation which can serve the purpose of a signature of two-dimensionalisation if we treat equation (\ref{ps}) anisotropically as follows.
\\
\indent Defining $\vec{l}\equiv\vec{x'}-\vec{x}$ and $\pa_{l_i}\equiv\na_i=\pa'_i=-\pa_i$, one can manipulate the equation (\ref{ps}) to get:
\be
\pa_t\la(\delta\theta)^2\ra+\na_i\la\delta v_i(\delta\theta)^2\ra=2\kappa\na_{ii}\la\theta^2\ra-4\kappa\la\na_i\theta\na_i\theta\ra-\epsilon_{ijk}\O_j\na_i\la l_k(\delta\theta)^2\ra
\label{psr}
\en
Now, owing to the anisotropy caused by rapid rotation, we may write $\la\delta\vec{v}(\delta\theta)^2\ra=\la\delta{v_{\bot}}(\delta\theta)^2\ra\vec{l}_{\bot}/{{l}_{\bot}}+\la\delta{v_{z}}(\delta\theta)^2\ra\vec{l}_{z}/{{l}_{z}}$ and as $\la(\delta\theta)^2\ra$ is proportional to terms quadratic in $l_{\bot}$ and $l_z$, in the limit $\kappa\rightarrow0$ and small scales, one can easily reach at the following relation:
\be
\la\delta v_{\bot}(\delta\theta)^2\ra|_{l_{\bot}=0}=0
\label{psc}
\en
This relation predicts that in the presence of rapid rotation, and hence anisotropy, on the small line segment parallel to axis of rotation the correlation in the L.H.S. of (\ref{psc}) vanishes. This may be readily used in numerics to check if the two-dimensionalisation has been achieved and hence may be treated as a signature of the effect.
\section{GOY turbulence}
In this section, we shall use GOY shell model\cite{Gledzer, Ohkitani} (modified appropriately) to investigate the behaviour of the structure functions and, thus, the signatures of two-dimensionalisation effect. 
One may ask immediately why one needs another shell model though Hattori {\it et. al.}\cite{Hattori} have already proposed a shell model --- an improved version of shell model by L'vov --- couples of years back.
To answer this question, let us collect the main results of that model: i) the exponent of the energy spectrum in the inertial range changes from $-5/3$ to $-2$, ii) no inverse cascade is detected with the increase in rotation rate, and iii) the PDF's of the longitudinal velocity difference doesn't match with the experiments.
Well, this field of studying the two-dimensionalisation effect is growing rapidly.
It has been confirmed well beyond doubt that the exponent overshoots the value $-2$ quite comfortably.
One may refer to the experiments by Morize {\it et. al.}\cite{Morize1}.
The model justifies its results by invoking weak-wave-turbulence-theory in which inverse cascade is not really shown.
This theory is a highly successful theory but one must be open-minded while dealing with problems as complex as turbulence and therefore, should take the experimental results at their face value.
That some experiments and numerics do show inverse cascade with increase in the rotation rate should motivate one to construct shell models that can mimic this effect.
As mentioned above, Hattori {\it et. al.}'s model finds PDF which mismatches with experiments and also, it requires a fluctuating part in the rotation rate to arrive at various results while in experiments and numerics there's no such part.
This again should make it clear that why at all we need another model.
Moreover, the numerical experiments done here are for unforced turbulence whereas Hattori {\it et. al.}'s model dealt with forced turbulence.
Hence, with due respect to the Hattori {\it et. al.}'s work, in this section we have tried to look at other possible shell model that can mimic the signatures of the two-dimensionalisation effect more closely.
\subsection{The model}
We have adopted the following strategy\cite{SSR1,SSR2} for the numerical experiments.
A specific form of GOY shell model for non-rotating decaying 3D turbulence is:
\be
\left[\f{d}{dt}+\nu k_n^2\right]u_n=ik_n\left[u_{n+2}u_{n+1}-\f{1}{4}u_{n+1}u_{n-1}-\f{1}{8}u_{n-1}u_{n-2}\right]^*
\label{revgoy1}
\en
This may be thought as a time evolution equation for complex scalar shell velocities $u_n(k_n)$ that depends on $k_n$ --- the scalar wavevectors labeling a logarithmic discretised Fourier space ($k_n=k_02^n$).
We choose: $k_0=1/6,\nu=10^{-7} \textrm{ and } n=1 \textrm{ to }22$.
The initial condition imposed is: $u_n=k^{1/2}e^{i\theta_n}$ for $n=1,2$ and $u_n=k^{1/2}e^{-k_n^2}e^{i\theta_n}$ for $n=3 \textrm{ to }22$ where $\theta_n\in[0,2\pi]$ is a random phase angle.
The boundary conditions are: $u_n=0$ for $n<1$ and $n>22$.
In the inviscid limit ($\nu\rightarrow 0$), equation (\ref{revgoy1}) owns two conserved quantities {\it viz.,} $\sum_n|u_n|^2$ (energy) and $\sum_n(-1)^nk_n|u_n|^2$ (helicity).
If the fluid is rotating then one may modify equation (\ref{revgoy1}) by adding a term $R_n=-i\left[\omega+(-1)^nh\right]u_n$ in the R.H.S.
$\omega$ and $h$ are real numbers.
It may be noted that this term, as is customary of Coriolis force, wouldn't add up to the energy.
The $(-1)^nh$ term part in $R_n$ has been introduced\cite{Reshetnyak} to have non-zero mean level of helicity that otherwise has a stochastic temporal behaviour and zero mean level.
Therefore, the appropriate shell model for rotating 3D turbulent fluid is:
\be
\left[\f{d}{dt}+\nu k_n^2\right]u_n=ik_n\left[u_{n+2}u_{n+1}-\f{1}{4}u_{n+1}u_{n-1}-\f{1}{8}u_{n-1}u_{n-2}\right]^*-i\left[\omega+(-1)^nh\right]u_n\label{revgoy2}
\en
We fix $h=0.1$ in our numerical experiments and test for $\omega=0.01,0.1,1.0$ and $10.0$.
We shall henceforth refer $\omega$ as rotation strength.
Numerical results are obtained using 500 independent initial conditions and 40 different statistically independent runs.
Inertial range has been taken as $n=4 \textrm{ to }15$ --- the range we are interested in.
We have, by the by, adopted slaved second order Adam-Bashforth scheme\cite{Pisarenko} to integrate equations (\ref{revgoy1}) and (\ref{revgoy2}).
\\
The $p$th order equal time structure function for the model has been defined as:
\be
\Sigma_p(k_n)\equiv\left\la\left|\textrm{Im}\left[u_{n+1}u_{n}\left(u_{n+2}-\f{1}{4}u_{n-1}\right)\right]\right|^{\f{p}{3}}\right\ra\sim k_n^{-\zeta_p}
\en
Such has been done to avoid possible existence of period three oscillations\cite{Kadanoff}.
The energy spectrum has been defined as: $E(k_n)=\Sigma_p(k_n)/k_n\sim k_n^{-m}$.
The mean rate of dissipation of energy is, of course, $\varepsilon=\la \sum_n\nu k_n^2 |u_n|^2\ra$ and flux through $n$th shell is calculated using the relation:
\be
&&\Pi_n\equiv\left\la-\f{d}{dt}\sum_{i=1}^{n}|u_i|^2\right\ra\\
\Rightarrow&&\Pi_n=\left\la-\textrm{Im}\left[k_nu_{n+1}u_{n}\left(u_{n+2}+\f{1}{4}u_{n-1}\right)\right]\right\ra
\en
For studying relative structure function scaling, the ESS scaling exponents\cite{Benzi} are taken as $\zeta_p^*\equiv\zeta_p/\zeta_3$.
$m$, $\zeta_p$ and $\zeta_p^*$ have all been calculated for inertial ranges only.
\subsection{The results}
The results are illuminating.
(Detailed discussion and figures are reported elsewhere\cite{SagarE3}).
One of them shows that as the rotation strength increases, the energy spectrum becomes steeper and the slope monotonically rushes from a value $\sim -5/3$ to a value of $\sim -7/3$; hence validating one of the two-dimensionalisation effect's signatures.
As we investigate into the direction of the flux in the inertial range regime, we can find that with the increase in rotation strength, first the forward cascade rate starts decreasing and then instances appear when at certain shells the flux direction reverses.
Again, the number of such shells increase as the rotation strength is enhanced; clearly suggesting that depletion in the rate of forward energy cascade.
Thus, yet another signature of two dimensionalisation has been upheld by the shell model.
At this point, it must be appreciated how important the inclusion of term $-i(-1)^nh$ in equation (\ref{revgoy2}) is in getting the effect of depletion in the rate of forward cascade.
By setting mean level of helicity above zero, it is this very term that --- in accordance with the arguments\cite{SagarE1} that it is the helicity that is causing this signature of two dimensionalisation effect to show up --- has empowered the model with the capacity to mimic the effect.
Attempts to get this very effect by setting $h=0$ have failed miserably in our numerical experiments.
The study of ESS in the shell model has been equally revealing.
It has been noted that the increase in the rotation strength is accompanied by a departure from the usual She-Leveque scaling\cite{She}.
But, the fact that at higher $p$, $\zeta_p$ seemingly becomes parallel to $p/2$ vs. $p$, is worth paying attention: This is in accordance with the direct numerical simulation (DNS) results\cite{Muller} and experimental results\cite{Baroud1}.
However, most interesting observation would be that, within the statistical error, $\zeta_p^*$ obtained for the rotating system via ESS coincides with that for the non-rotating ones.
Probably, this extends the ESS for 3D fluids even further by implying that rotation keeps ESS scaling intact, even though usual $\zeta_p$ changes owing to rotation.
Of course, only experiments and DNS can judge if this really is true for real fluid turbulence: GOY shell, after all, is just a model that remarkably reproduces many characteristic features of turbulence by only using a fraction of the computation power needed by DNS.
In this context, one might be well aware that some modified versions of GOY model invented to model the distinguishing features of 2D turbulence have been shown to be rather useless\cite{Aurell}.
One, thus, always has to be careful while dealing with simplified models of turbulence.
\section{Discussions and Conclusions}
The Kolmogorov-Landau approach has been invoked in 2D homogeneous isotropic unforced fluid turbulence to arrive at the various correlation functions earlier obtained using different methods.
Also, some experimentally verifiable correlation functions in the dissipation range have been derived.
The results derived here are `exact' (though not rigorous) something which is a far cry in the literature on turbulence.
However, we have been careless enough to assume the existence of $\eta$ when $\nu\rightarrow 0$.
If $\eta$ doesn't exist, the one-eighth law is in jeopardy.
It is really unfortunate for the law that it has been rigourously proved\cite{Eyink2,Tran} that enstrophy dissipation is not possible for any 2D Euler solutions with finite enstrophy.
Thus, $\eta$ may exist in the inviscid limit only when one takes rather ill-defined initial conditions for which the total initial enstrophy is infinite.
In view of this, one must not be surprised at all if numerics and experiments fail to uphold the one-eighth law in many a situation.
This very law of 2D turbulence, therefore, doesn't enjoy the same classic status as the Kolmogorov law of 3D turbulence.
\\
Studies with structure functions of QG turbulence have again showcased how handy and useful the Kolmogorov-Landau approach can prove to be.
The results (\ref{45}) and (\ref{energy_cascade_S3qg}) naturally agree with what has been arrived at by Lindborg\cite{Lindborg} earlier.
Within the domain of the approximations made these results are exact, something worth getting as the literature of turbulence is comparatively barren as far as exact relations are concerned.
However, the hypothesis of the equipartition of energy used in equation (\ref{derteqg}) is as questionable as the assumption of isotropy in the sense of Charney.
This hypothesis needs to be put on more firm basis.
Again, the existence of $\varepsilon_q$, like $\eta$, is questionable when $\nu\rightarrow 0$.
However, studies in 2D and QG turbulences in the perspective of the velocity structure functions hint that finding the velocity structure functions in the rotating flows with a view to unfolding the mysteries of two-dimensionalisation effect might not be just a wild goose chase.
\\
We emphasis on the fact that the form of two point third order structure function in a slowly rotating homogeneous 3D turbulence can strongly indicate the initiation of the effect of two-dimensionalisation of 3D turbulence.
It barely needs to be mentioned that the relations are quite interesting and pertinent (at least within the approximations made in the calculations) -- something which, as is being said again and again, is worth getting in the literature of turbulence since exact relations are very few therein.
So any theory developed in the limit of $Ro\rightarrow 0$ and $Re\rightarrow \infty$, must satisfy the relation (\ref{rev25}) derived in this article in the limit of low $\O$ or explicitly violate the assumptions made to arrive at the result; in this sense the relation may prove to be of high importance.
Moreover, true reason behind the so called two-dimensionalisation of turbulence has been figured out which accounts for the correct energy cascade direction and the correct energy spectrum found in the experiments and simulations.
To settle the problem more neatly, it has been proposed that the study of passive scalars in rotating turbulence may prove to be of benefit.
\\
Shell models have been successfully used to study statistical properties of turbulence of many authors (see ref-(\cite{Bohr}) and ref-(\cite{Biferale}) for details).
Most of them dealt with the homogeneous isotropic turbulence.
Hattori {\it et. al.} gave a shell model for rotating turbulence.
Their result could be improved in the face of the rapid progress in the field.
We, thus, have used a modified version of GOY shell model to study the two-dimensionalisation effect.
Some results of the model are, no doubt, consistent with experiments and DNS.
However, one can always question the effectiveness of the signatures discussed herein because $i$) a scaling law for a single-component spectrum, though heavily used in literature, has poor meaning in the strongly anisotropic configuration relevant to pass from 3D-2D; different power laws can be found in terms of $k_z$, $k_{\bot}$ and $k$ in contrast to the 3D isotropic case, and $ii$) the inertial wave-turbulence theory is not consistent with an inverse cascade.
Actually in weak-wave turbulence, getting rid provisionally of helicity and polarisation spectra, a two-component energy spectrum $e(k,\cos\theta)$ with $\cos\theta=k_z/\sqrt{(k^2_z+k^2_{\bot}})$ is found to be useful; if $E$ denotes the traditional spherically averaged spectrum, the anisotropic structure is one of the best ways to quantify all intermediate states from isotropic 3D (with $e=E(k)/(4\pi k^2)$) to 2D state (with $e=E(k_{\bot})/(2\pi k_{\bot})\delta(k_z)$).
Two-dimensional trend is therefore linked to a preferred concentration of spectral energy towards the transverse wave-plane $k_z=0$.
This concentration, however, does not necessarily yield an inverse cascade.
A reasonable suggestion, in the light of this discussion, would be that in the shell model for rotating turbulence $k$ should be interpreted as $k_{\bot}$.
It may be concluded that this study has put the equation (\ref{revgoy2}) as a very firm shell model for the rotating 3D turbulent flows; after all, it explains the observed signatures of the two-dimensionalisation effect so closely.
Probably, this model discussed in this article and the model due to Hattori {\it et. al.} can together yield a much better shell model for the rotating turbulence.
\\
In the absence of solutions of the Navier-Stokes equations, these various results regarding two-dimensionalisation can only be checked by data from experiments, and this endeavour seems to be difficult at present.
So in the closing, we hope that validity of the results derived will be checked both numerically and experimentally in near future.
After all, the structure functions traditionally provide checks for any plausible theory of turbulence.
\acknowledgements
The author thanks his supervisors Prof. J.K. Bhattacharjee and Dr. Partha Guha for fruitful suggestions; and also his friends Anjan, Ashish, Ayan, Prasad, Rudra, Samriddhi and Subhro for their support in various forms.
The author is indebted to Prof. Uriel Frisch and Prof. E. Lindborg for helpful correspondences.
CSIR (India) is gratefully acknowledged for supporting him financially through a senior research fellowship.
This article is dedicated to Prof. K. Srinivasan who has enthused the author into serious scientific research.
%%%%%%%%%%%%%%%%%%%%%%%%%%%%%%%%%%%%%%%%%%%%%%%%%%%%%%%%%%%%%%%%%%%%%%%%%%%%%%%%%%%%%%%%%%%%%%%%%%%%%%%%%%%%%%%%%%%%%%%%%%%%
%

%%%%%%%%%%%%%%%%%%%%%%%%%%%%%%%%%%%%%%%%%%%%%%%%%%%%%%%%%%%%%%%%%%%%%%%%%%%%%%%%%%%%%%%%%%%%%%%%%%%%%%%%%%%%%%%%%%%%%%%%%%%%
%
\end{document}